\documentclass[paper,notoc]{JHEP3}
\usepackage{epsfig,cite}
\usepackage{amsbsy} 
\usepackage{epsfig}
\usepackage{graphicx}
\usepackage{amssymb,amsfonts,amsmath,bbold}

\def \ep{\epsilon}

\title{\boldmath The one-loop gluon amplitude for heavy-quark production
at NNLO}

\author{Charalampos Anastasiou\\
  Institute for Theoretical Physics, ETH Zurich,\\
  8093 Zurich, Switzerland\\
  E-mail: \email{babis@phys.ethz.ch}}
\author{S. Mert Aybat\\
  Institute for Theoretical Physics, ETH Zurich,\\
  8093 Zurich, Switzerland\\
  E-mail: \email{aybat@phys.ethz.ch}}

\abstract{ 
We compute the one-loop QCD amplitude for the process 
$gg \to Q\bar{Q}$ in dimensional regularization through 
order $\epsilon^2$ in the dimensional regulator and for 
arbitrary quark mass values.  
This result is an ingredient of the 
NNLO cross-section for heavy quark production at hadron 
colliders. The calculation is performed in conventional 
dimensional regularization, using well known reduction 
techniques as well as a method based on recent ideas  for 
the functional form of one-loop integrands in four dimensions. 
} 
\keywords{NLO and NNLO computations}

\begin{document}

\section{Introduction}
\label{sec:introduction}

Heavy quarks are related to some of the most exciting physics studies  at
hadron colliders.  It is very likely that the  third quark family  has
a special role in the breaking of electroweak symmetry.  
The mass of the top quark measured at the Tevatron \cite{Group:2008vn}
is a sensitive probe of the theory for the breaking 
of the electroweak symmetry,  and  
it constrains, for example, together with other electroweak precision measurements 
the mass of the Higgs boson. 

So far, top quarks have only been  produced at the Tevatron. 
Detailed studies  of the properties of the top
quark will be  a  main theme for the experiments at the Large Hadron
Collider. 
The LHC is often termed a ``top factory'' 
since it is capable of producing many such particles per second. The 
top-pair production cross-sections will be measured with a negligible 
statistical uncertainty in comparison to the most optimistic
predictions for the attainable accuracy of  theoretical  calculations. Systematic experimental uncertainties could be
nevertheless sizeable.  For  example, the CMS collaboration
anticipates to measure the top quark cross-section with an early systematic
uncertainty of $10\%$ to $20\%$ depending on the  decay channel of the 
top quarks \cite{CMS}.  These systematic errors may be further
improved with a large integrated luminosity. 

A large body of work has been devoted to obtaining  precise
theoretical estimates for heavy quark  cross-sections. Next-to-leading-order 
(NLO) QCD corrections for the spin-averaged cross-section have been 
computed in Refs.~\cite{Nason:1987xz,Beenakker:1988bq,Mangano:1991jk}. 
Calculations with the full spin-dependence of the heavy quark decays  
were performed in Refs.~\cite{Bernreuther:2001bx,Bernreuther:2004jv}. 
The effect of soft-gluon resummation at the leading and
next-to-leading logarithmic approximations was accounted for in
Refs.~\cite{Laenen:1993xr,Berger:1996ad,Kidonakis:1997gm,Bonciani:1998vc}. Soft gluon resummation effects beyond the next-to-leading logarithmic approximation were included in Refs.~\cite{Kidonakis:2000ui}.  
NLO QCD calculations and parton shower event generators have been matched using the MC@NLO approach in Ref.~\cite{Frixione:2003ei}.

A recent theoretical estimate for the total cross-section at 
NLO in QCD at the
LHC is \cite{Cacciari:2008zb}:  
\begin{equation*}
\sigma^{\rm NLO}_{t\bar t}({\rm LHC},m_{\rm top}=171 {\rm GeV})=875^{+102(11.6\%)}_{-100(11.5\%)}({\rm scales})^{+30(3.4\%)}_{-29(3.3\%)}({\rm PDFs})\,{\rm pb}\,.
\end{equation*}
 Similar analyses have been performed in Refs.~\cite{Moch:2008qy}. The above theoretical uncertainty is marginally  sufficient
 for  a comparison with the projected systematic experimental
 errors. It will be important to improve upon it by performing a complete NNLO calculation.

The only  two cross-sections for hadron collider processes which have 
been computed beyond NLO in QCD are Drell-Yan lepton-pair production \cite{Hamberg:1990np,Harlander:2002wh,Anastasiou:2003yy,Melnikov:2006di}
and Higgs boson production \cite{Harlander:2002wh,Anastasiou:2002yz,Ravindran:2003um,Anastasiou:2004xq,Anastasiou:2005qj,Catani:2007vq,Anastasiou:2007mz,Grazzini:2008tf}.  It is interesting that while in Drell-Yan
production the NLO  theoretical estimate from varying the renormalization
scales turns out to be reliable, this is not the case in the gluon
initiated process of Higgs boson production. It has also been
observed that the NLO calculations for Higgs production fail for efficiencies when experimental cuts vetoing radiation are applied \cite{Anastasiou:2008ik}. Similar cuts must be applied in various analyses
(e.g. \cite{DavatzCMSnote}) for top-production when this is a background process
for other interesting signals. 

While there are many commons among top-pair production and Higgs boson 
production, such as the dominant contribution from gluonic initial
states and a heavy invariant mass being
produced in the final state, there are good signs  that the top-pair
cross-section may have  a better converging perturbative expansion;
for example  the NLO K-factor \cite{Campbell:2006wx} is smaller for top production
than Higgs production. It is also very encouraging that event
generators with NLO matching such as MC@NLO are in very good agreement with NNLO \cite{Anastasiou:2008ik} in estimating experimental efficiencies for Higgs boson production.   
Nevertheless,  it is necessary that any good intuition about  
higher order effects beyond NLO is verified  explicitly
with an NNLO calculation.

There has been a  significant recent effort towards computing the NNLO
corrections.  This is a computation which requires  several
ingredients. The virtual 2-loop QCD corrections in the limit $s,|t|,|u| \gg m_Q^2$ were computed in 
Refs.~\cite{Czakon:2007ej,Czakon:2007wk} using a 
factorization theorem \cite{Mitov:2006xs} and
explicit two-loop computations for massless parton scattering \cite{Anastasiou:2000kg}. 
The full two-loop amplitude for $q\bar q\to Q\bar Q$ was computed numerically in Ref.~\cite{Czakon:2008zk}. 
The corresponding analytic result for the sub-amplitude 
with fermionic one-loop insertions in propagators was computed  in
Ref.~\cite{Bonciani:2008az}.

A lot more work is  required for a  complete NNLO cross-section. For
example, the NNLO calculation should include the computation for 
the cross-section $pp \to t \bar{t} \mbox{jet} + X$ at NLO,  which  has 
been recently computed in Ref.~\cite{Dittmaier:2007wz}. However, at NNLO there exist additional infrared singularities arising from  one or two massless partons being unresolved. For the
result of Ref.~\cite{Dittmaier:2007wz} to be utilized, a new subtraction formalism for
NNLO calculations is required, 
extending non-trivially the theoretical knowledge gained in 
the NNLO computations for $e^+e^- \to 3jets$ \cite{GehrmannDeRidder:2007jk}
and $pp \to H +X$ \cite{Catani:2007vq}. An alternative approach would be  
to apply the methods followed in Refs.~\cite{Melnikov:2006di,Anastasiou:2004xq,Anastasiou:2004qd} used for Higgs
boson  and Drell-Yan production. In both approaches there
are several new problems which need to be addressed, and success is by no
means guaranteed.

An important ingredient of the NNLO computation is the one-loop
amplitudes for $gg \to t \bar{t}g$, $gg \to t\bar{t}$ and the likes
replacing gluons with light quarks in the  external states. In the
absence of  a subtraction method, these amplitudes need to be computed 
through ${\cal O}(\epsilon^2)$ in the dimensional regulator $\epsilon
= 2 - \frac{d}{2}$. These amplitudes have been computed to sufficiently high order in $\ep$ in Ref.~\cite{Korner:2002hy}. The one-loop squared contribution for $q\bar q \to t \bar{t}$ was recently 
given in Ref.~\cite{Korner:2008bn}.

In this paper we compute the gluonic
one-loop amplitude $gg \to \bar{t} t$ using fully numerical and analytical methods. This is a rather humble
computation in comparison to the remaining problems for obtaining 
a NNLO cross-section. However, many of the problems which are foreseen
for later tasks, especially
related to the numerical efficiency of computing the one-loop
contributions for sub-processes with five external legs in the
presence of additional phase-space singularities, may benefit
from an efficient evaluation of the loop amplitudes.

Our goal in this paper is to achieve an efficient evaluation of the squared one-loop amplitude for $gg\to Q\bar Q$. Recently, powerful new methods \cite{Ossola:2006us,Ossola:2007bb,Ossola:2008xq,delAguila:2004nf,Ellis:2007br,Giele:2008ve,Berger:2008sj,Ellis:2007qk,Giele:2008bc} 
have emerged for the computation of one-loop amplitudes 
based on the groundbreaking work of Refs.~\cite{Ossola:2006us,delAguila:2004nf}. These methods have started to make an impact in phenomenology \cite{Binoth:2008kt} but are still lacking the maturity of other approaches \cite{Bredenstein:2008zb,Lazopoulos:2008de,Giele:2004iy,Campbell:2007ev,Campanario:2008yg} which lead to
impressive applications of an extraordinary technical difficulty. 
Due to their conceptual simplicity, however, it is realistic to anticipate
a significant impact with the new methods. 

We have  performed two independent computations of the one-loop $gg
\to t \bar{t}$ amplitude through order ${\cal O}(\epsilon^2)$. 
First, mostly for testing purposes, we use the method 
of Davydychev \cite{Davydychev:1991va,Anastasiou:1999bn} which is 
analytic. Our main computational method is a new extension of the 
the method proposed in Ref.~\cite{Anastasiou:2006jv}, which was  also 
analytic. Here we show how to achieve a numerical implementation.
  
Ref.~\cite{Anastasiou:2006jv}  proposed to perform reductions of
tensor integrals in two stages: first, a reduction in exactly four dimensions where 
every propagator is modified by adding a common mass parameter, 
and second, a mapping of the results in the four dimensional reduction
to a reduction in $D=4-2\epsilon$ dimensions.  It was proposed then
that  the four dimensional reduction could be performed using the 
method of ~\cite{Ossola:2006us}.  For the mapping to work, the
analytic behavior in the mass parameter was required.  Here, we find
that  a numerical reduction for a fixed number of  values of the mass 
parameter is sufficient to reconstruct the  full dimensional dependence of the
master integral coefficients at each phase-space point.

A very elegant solution to the problem of a D-dimensional
reduction for one-loop amplitudes directly and not just tensor integrals 
was given in Ref.~\cite{Giele:2008ve}. In fact, the $gg \to t \bar{t} g$
one-loop amplitude which is needed for the NNLO computation 
of top-pair cross-section was computed recently in
Ref.~\cite{Ellis:2008ir}. The method of Ref.~\cite{Giele:2008ve} capitalizes on unitarity
ideas~\cite{unitarity,Bern:1994zx,Bern:1995db,Bern:1996,Bern:1997sc,BCFgeneralized} 
and has the advantage of being able  to use  tree amplitudes for the
reconstruction  of master integral coefficients.

Unitarity methods require the evaluation of
one-loop amplitudes in renormalization schemes such as the four
dimensional helicity scheme (FDH) \cite{Bern:2002zk} where external particles are in 
four dimensions.  The FDH scheme has been successfully used for the 
evaluation of one-loop and two-loop amplitudes in QCD. However, it is
known to be  inconsistent for the two-loop  amplitude $gg \to h$ \cite{Anastasiou:2008rm}, and 
at higher loop orders the renormalization of the strong
coupling is not equivalent to the conventional Dimensional Reduction scheme \cite{Harlander:2006rj}. It is not yet clear to us whether the FDH scheme is  a
consistent scheme at NNLO for top-pair production. Our method allows
us to compute the required one-loop amplitudes in conventional dimensional regularization (CDR) \cite{'tHooft:1972fi} by using a projection method in order to extract the spin dependence of the amplitude, similar to what it was used for in Ref.~\cite{Garland:2002ak}.

The paper is organized as follows: in section~\ref{sec:notation} we introduce
our notation. In section~\ref{sec:method} we discuss the analytic method we
used to check our results and our new numerical method. In
section~\ref{sec:results} we present our results for the $gg\to Q\bar Q$ amplitude. Finally
we conclude with section~\ref{sec:conclusions}.

\section{Notation}
\label{sec:notation}

\begin{figure}[h]
\begin{center}
\includegraphics[width=6.5cm]{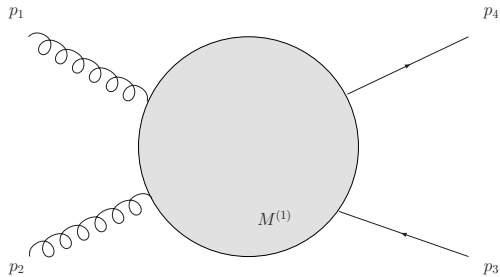}
\caption{Gluon amplitude for heavy quark production}
\label{amplitude}
\end{center}
\end{figure}
We consider the scattering process 
\begin{equation}
g(p_1; a_1) + g(p_2; a_2) + Q(p_3; \, i_3) + \bar{Q}(p_4;\, i_4) \to 0
\end{equation}
where we take all momenta $p_i$ incoming so that 
\begin{equation}
p_1+p_2+p_3+p_4=0.
\end{equation}
We denote by $a_1,a_2 = 1 \ldots N^2-1$ and by $i_3,i_4=1\ldots N$ the 
color of the external gluons and external quarks correspondingly. 
In addition, the momenta satisfy
\begin{equation}
p_1^2=p_2^2=0,\qquad  p_3^2=p_4^2=m_Q^2,
\end{equation}
where we apply the approximation that the 
heavy quarks are on-shell and $m_Q$ denotes 
the heavy quark mass. We define the Mandelstam variables,
\begin{equation}
s \equiv (p_1+p_2)^2, \qquad t \equiv (p_2+p_3)^2 \quad \mbox{and}  \quad 
u \equiv (p_1+p_3)^2;
\end{equation}
they satisfy
\begin{equation}
s+t+u = 2 m_Q^2.
\end{equation}

We present here the unrenormalized QCD  amplitude with $N_l=0$ light quark 
flavors and $N_h=0$ heavy flavors different than the flavor of the produced 
quark pair. 
In our main result, we also drop diagrams with self-energy corrections 
on the external legs, envisaging an on-shell renormalization procedure 
for the heavy quark mass and a decoupling of the heavy flavor from the 
renormalization group evolution of the strong coupling. 
The general result for the unrenromalized amplitude, having  $N_h,N_l \neq 0$ 
will be given  in Section~\ref{sec:results}. 
Infrared and ultraviolet singularities are regularized 
using conventional dimensional regularization 
performing both the $\gamma$-matrix algebra and loop integrations in 
$D = 4-2\epsilon$ dimensions \cite{'tHooft:1972fi}.

We write the amplitude as a perturbative series in 
the bare coupling  $\alpha_s$, 
\begin{equation}
\label{amplitude}
M_{gg\to Q\bar Q}=4\pi\alpha_s\left[M_{gg\to Q\bar Q}^{(0)}+\frac{\alpha_s}{4\pi}\,\Gamma(1+\ep)\,\left(\frac{4\pi\mu^2}{m_Q^2}\right)^{\ep}\,M_{gg\to Q\bar Q}^{(1)}\right]
+\mathcal{O}(\alpha_s^3). 
\end{equation} 
The tree and one-loop amplitudes are expressed in terms of the color
functions
\begin{equation}
\label{color_basis}
c_1 = T^{a_2}_{i_4j} T^{a_1}_{ji_3},\quad 
c_2 = T^{a_1}_{i_4j} T^{a_2}_{ji_3},
\quad 
c_3 = \delta^{a_1a_2} \delta_{i_4i_3},
\end{equation} 
as
\begin{eqnarray}
M_{gg\to Q\bar Q}^{(n)} &=& \sum_{L=1}^3\mathcal{M}_L^{(n)}\,c_L\,,
\end{eqnarray}
where $n=0,1$. The  $SU(N)$ generators in the adjoint representation 
are normalized as $tr\left( T^a T^b\right) = \frac{1}{2}\delta^{ab}$.

We present the amplitude for a specific choice of gluon polarizations, 
requiring that the polarization vector of one gluon is perpendicular
to the momentum of the other gluon, 
\begin{equation}
\ep_1 \cdot p_2 = \ep_2 \cdot p_1 =0. 
\end{equation}
With this constraint on gluon polarizations the color coefficients
$\mathcal{M}_L$ can be expressed in  terms of a basis of ``spin structures'' as
\begin{equation}
\label{amplitude_sspace}
\mathcal{M}_L^{(n)} = \sum_i  d_{i,L}^{(n)}  \Gamma_i,
\end{equation}
where 
\begin{align}\label{gg_spin}
\Gamma_1 &= m_Q\,\ep_1\cdot \ep_2\,\bar v(p_4)\,u(p_3)\, &\Gamma_2 &= m_Q\,\ep_1\cdot p_3\,\ep_2\cdot p_3\,\bar v(p_4)\,u(p_3)\nonumber\\
 \Gamma_3 &= \ep_1\cdot \ep_2\,\bar v(p_4) \not \! p_1\, u(p_3)\, &\Gamma_4 &= \ep_1\cdot p_3\,\ep_2\cdot p_3\,\bar v(p_4) \not\! p_1\,u(p_3) \nonumber\\
\Gamma_5 &= \ep_2\cdot p_3\,\bar v(p_4)\not\! \ep_1\,u(p_3)\,&\Gamma_6 &= \ep_1\cdot p_3\,\bar v(p_4)\not\! \ep_2\,u(p_3)\nonumber\\
\Gamma_7 &= \ep_2\cdot p_3\,\bar v(p_4)\not\! \ep_1\not\! p_1\,u(p_3)\,&\Gamma_8 &=
\ep_1\cdot p_3\,\bar v(p_4)\not\! \ep_2 \not\! p_1\,u(p_3)\nonumber\\
\Gamma_9 &= m_Q\,\bar v(p_4)\not\! \ep_1\not\! \ep_2\,u(p_3)\,&\Gamma_{10} &= m_Q\,\bar v(p_4)\not\! \ep_1\not\! p_1\not\! \ep_2\,u(p_3).
\end{align}

In terms of these spin structures and the color basis of
Eq.~(\ref{color_basis}), the born amplitude is given by
\begin{eqnarray}
i\,\mathcal{M}_1^{(0)} &=& -\frac{2}{s}\,\Gamma_3 -
\frac{2}{u-m_Q^2}\,\Gamma_3 +
\frac{2}{u-m_Q^2}\,\Gamma_6 -
\frac{1}{u-m_Q^2}\frac{1}{m_Q}\Gamma_{10}\,\nonumber\\
i\,\mathcal{M}_2^{(0)} &=& \frac{2}{s}\,\Gamma_3 +
\frac{2}{t-m_Q^2}\,\Gamma_6 - \frac{1}{t-m_Q^2}\frac{1}{m_Q}\Gamma_{10}\,,\nonumber\\
i\,\mathcal{M}_3^{(0)} &=& 0\,. 
\end{eqnarray}

In this  paper we present the one-loop 
coefficients $d_{i,L}^{(n)}$ through order ${\cal O}(\epsilon^2)$, 
which we compute with both an analytic and a purely numerical method. 
The numerical method relies on extracting these coefficients by 
acting with  a set of projectors on the  one-loop amplitude.

Let us multiply both sides of Eq.~(\ref{amplitude_sspace}) by
$\Gamma_j^{\dagger}$ and sum over gluon and quark polarizations 
\begin{eqnarray}
\sum_{\rm polarizations}\mathcal{M}_L^{(n)}
\Gamma_j^{\dagger} &=& \sum_i \sum_{\rm polarizations}\, d_{i,L}^{(n)} \Gamma_i\,\Gamma_j^{\dagger} \nonumber\\
&=& \sum_i \, d_{i,L}^{(n)} \Gamma_{ij}, \nonumber\\ 
\end{eqnarray}
where we defined the symmetric matrix $\Gamma_{ij}$ to be
\begin{equation}\label{gamma}
\Gamma_{ij}\equiv\sum_{\rm polarizations}\Gamma_i\,\Gamma_j^{\dagger}\,.
\end{equation}
The coefficients $d_i$ are then given by
\begin{equation}
\label{d_i}
d_{i,L}^{(n)} =\sum_j\,\left(\Gamma^{-1}\right)_{ji}\sum_{\rm polarizations} 
\mathcal{M}_L^{(n)} \Gamma_j^\dagger.
\end{equation}
The matrix $\Gamma_{ij}$ and the inverse matrix $\Gamma^{-1}_{ij}$ are
evaluated in CDR and  we list their components in Appendix~\ref{matrices}.

Once we know the coefficients $d_i$ using Eq.~(\ref{d_i}), we can
calculate the one-loop amplitude squared $|M_{gg\to Q\bar Q}|^2$
summed over external quark and gluon polarizations and averaged over initial state color and polarizations to be
\begin{eqnarray}
\sum_{polarizations}\overline{|M_{gg\to Q\bar Q}^{(1)}|^2} &=&  \alpha_s^4\,\frac{1}{256 (1-\ep)^2}\,\Gamma(1+\ep)\left(\frac{4\pi\mu^2}{m_Q^2}\right)^{\ep}\nonumber\\
&\ & 
\hspace{-30mm}\times \sum_{L,M=1}^3\left[\sum_i\,\sum_j\,d_{i,L}(\epsilon,\mu;m_Q,\{p_i\})\,d_{j,M}^*(\epsilon,\mu;m_Q,\{p_i\})\,\Gamma_{ij}\right]\,C_{LM}
\label{eq:square}
\end{eqnarray}
where we average over color and polarizations for the  external
gluons. 
We have also defined the color matrix $C_{LM}\equiv tr\left(c_Lc_M\right)$, 
given by
\begin{equation}
C_{LM} = \left (
\begin{array}{ccc}
N\,C_F^2 & -\frac{1}{2}\,C_F & N\,C_F \\ \\
-\frac{1}{2}\,C_F & N\,C_F^2 & N\,C_F \\ \\
N\,C_F & N\,C_F& 4N
\end{array}
\right)\,,  
\end{equation}
with $C_F=\frac{N^2-1}{2N}$.
 
We expand the one-loop amplitude squared in $\epsilon$,
\begin{equation}
\label{A}
\sum_{polarizations}\overline{|M_{gg\to Q\bar Q}^{(1)}|^2} 
=\alpha_s^4\,\frac{\Gamma(1+\ep)^2}{(1-\ep)^2}\left(\frac{4\pi\mu^2}{m_Q^2}\right)^{2\ep}\,\left[A_4\frac{1}{\ep^4}+A_3\frac{1}{\ep^3}+A_2\frac{1}{\ep^2}+A_1\frac{1}{\ep}+A_0\right]\,+\mathcal{O}(\ep)\,.
\end{equation}
In Section~\ref{sec:results} we present plots for $A_4$, $A_3$, $A_2$, $A_1$ and
$A_0$  as a fuction of  $-t/s$ for fixed $s=16\,m_Q^2$ and $m_Q=1$. We
also provide analytic results. 

When the heavy quarks are an intermediate state and a decay 
is considered, we simply need to replace the matrix $\Gamma_{ij}$ with the
analogous matrix  for the specific decay in Eq.~(\ref{eq:square}).

\section{The Method}
\label{sec:method}
We have performed the calculation using two independent methods for
the reduction to master integrals.  Both methods are based on
Feynman diagrams, which we generate using QGRAF~\cite{Nogueira:1991ex}. 
Algebraic manipulations involving  color and Lorentz  indices are
performed using FORM~\cite{Vermaseren:2000nd}. 

\subsection{Analytic approach}
In our first approach we perform an explicit analytic evaluation of all tensor
integrals which appear in the one-loop amplitude.  Following the method
of Davydychev~\cite{Davydychev:1991va, Anastasiou:1999bn} we write 
tensor integrals in terms of scalar integrals where the denominators
are raised to integer powers of propagators   and the dimensionality of the integral is
increased by an even number. 
Schematically, 
\begin{eqnarray}
 \int d^Dk \frac{k^{\mu_1} \ldots k^{\mu_r}}{
\prod_j \left[ \left(k+q_j\right)^2 -m_j^2 \right] 
} &=& \sum_{{\cal T}} {\cal T}^{\mu_1 \ldots \mu_r}  \sum_{m} \sum_{\nu_1 \ldots \nu_n} 
\alpha_{m;\nu_1 \ldots \nu_n}^{{\cal T}} 
\nonumber \\
&& 
\times
\int d^{D+2m}k \frac{1}{
\prod_j \left[ \left(k+q_j\right)^2 -m_j^2 \right]^{1+\nu_j} }
\end{eqnarray}
where ${\cal T}^{\mu_1 \ldots \mu_r}$ are all tensors of rank $r$ 
that can be constructed from products of the metric tensor and the 
momenta $q_i$.  $m$ and $\nu_j$ are non-negative integers and 
the coefficients $\alpha_{m;\nu_1 \ldots \nu_n}^{{\cal T}} $ are also integers. 

As a next step we perform a reduction of the dimensionally shifted integrals with extra powers of denominators to master integrals in the same dimension. These are  
typically scalar integrals with unit powers of propagators. 
\[
\int d^{D+2m}k \frac{1}{
\prod_j \left[ \left(k+q_j\right)^2 -m_j^2 \right]^{1+\nu_j} } 
\longrightarrow
\int d^{D+2m}k \frac{1}{
\prod_j \left[ \left(k+q_j\right)^2 -m_j^2 \right]^{1} } 
\]
We perform this reduction using the program AIR~\cite{Anastasiou:2004vj}. 

Master integrals in $D+2m$ dimensions can be reduced in terms of 
master integrals in $D$ dimensions~\cite{Davydychev:1991va,Bern:1993kr,Campbell:1996zw}. 
\begin{equation}
\label{eq:dimshift}
\int d^{D+2m}k \frac{1}{
\prod_j \left[ \left(k+q_j\right)^2 -m_j^2 \right]^{1} } 
\longrightarrow
\int d^{D}k \frac{1}{
\prod_j \left[ \left(k+q_j\right)^2 -m_j^2 \right]^{1} }. 
\end{equation}
This concludes our analytic reduction of tensor integrals to master integrals. 
In Appendix \ref{app_IBP} we present generic dimensional shift identities with
arbitrary masses and arbitrary external momenta for box, triangle and bubble
topologies, which we obtained with the method outlined in this section.

This analytic method has worked fine for our application in this paper. 
However, we had to deal with rather large expressions in intermediate stages 
as well as in the final result. This can be neatly avoided with an alternative numerical method which we present here. We anticipate that this alternative approach will also facilitate the evaluation of the loop contributions to the NNLO cross-section from processes with five external legs.

\subsection{Numerical approach}

We consider the one-loop amplitude in $D=4-2\epsilon$ dimensions
\begin{equation}
\label{integral_D}
{\cal  M}(D=4-2\epsilon)=
\int
\frac{d^Dk}{i\pi^{D/2}}
\frac{{\cal A}(k)}{\prod_j \left[ \left( k+q_i\right)^2 -m_i^2 \right]},
\end{equation}
where $k$ is the $D$ dimensional loop-momentum.
Following \cite{Bern:1995db,Bern:1996}.  
we may decompose the $D$ dimensional integral into a $4$ and a $D-4$ dimensional integral by writing $k$ as 
\begin{equation}
\label{mom_decompose}
k=l+\mu\,,
\end{equation}
where $l$ is the $4$ dimensional part and $\mu$ is the $D-4$ dimensional part 
for the loop momentum. We have, 
\begin{equation}
\label{lsquared}
k^2=l^2-\mu^2.
\end{equation}
In addition, we consider all combinations of external momenta $q_i$ in four 
dimensions, so that
\[ 
\mu \cdot q_i=0.
\]
With this decomposition of the dimension and the loop-momentum, 
we can rewrite the integral in Eq.~(\ref{integral_D}) as
\begin{equation}
\label{integral_decomposed}
{\cal  M}(D=4-2\epsilon)=
\frac{-i}{\pi^2\Gamma(-\epsilon)}\int d\mu^2 
\left(\mu^2\right)^{-\epsilon-1}
\int d^4l
\frac{{\cal A}(l, \mu)}{\prod_j \left[ \left( l+q_i\right)^2 -\left( \mu^2+ m_i^2 \right) \right]}.
\end{equation}
In the above equation, the number of dimensions $D=4-2\epsilon$ for the original 
loop momentum $k$ is assumed to be bigger than four. Before we interpret the 
inner integration as an integration in four dimensions, the issue of the dimensionality of $\gamma$ matrices, as well as 
polarization vectors and spinors for the external states needs to 
be addressed. These are usually dealt with by performing the evaluation of the 
inner integrand using the FDH scheme \cite{Bern:2002zk}. As we explained in the introduction, 
we would like to avoid performing the computation in FDH.  

These issues do not arise when the amplitude is multiplied with the matrices 
$\Gamma_i$ of Section~\ref{sec:notation} and the trace is taken. This is sufficient in order to reconstruct the full
amplitude once the  matrix $\Gamma_{ij} =tr\left(\Gamma_i^\dagger
  \Gamma_j\right)$ and  its inverse are known (appendix~\ref{matrices}). 
We then write, 
\begin{equation}
\label{integral_decomposed_traced}
tr \left[ \Gamma_i^\dagger {\cal  M}(D=4-2\epsilon) \right]=
\frac{-i}{\pi^2\Gamma(-\epsilon)}\int d\mu^2 
\left(\mu^2\right)^{-\epsilon-1}
\int d^4l
\frac{
tr \left[ \Gamma_i^\dagger {\cal A}(l, \mu) \right]_{CDR}
}
{\prod_j \left[ \left( l+q_i\right)^2 -\left( \mu^2+ m_i^2 \right) \right]},
\end{equation}
where we have performed the trace in CDR before decomposing the loop momentum. 
Therefore, the  traced integrand depends explicitly on $\epsilon$ from the $\gamma$-matrix algebra in CDR. 
\begin{eqnarray}
\label{integral_decomposed_traced_cdr}
tr \left[ \Gamma_i^\dagger {\cal  M}(D=4-2\epsilon) \right]  &=&
\frac{-i}{\pi^2\Gamma(-\epsilon)}\int d\mu^2 
\left(\mu^2\right)^{-\epsilon-1} \nonumber \\
&& \hspace{-4.7cm}
\times
\int d^4l
\frac{
\sum_{n=0}^{3}\epsilon^n \left. tr \left[ \Gamma_i^\dagger {\cal A}(l, \mu) \right]_{CDR} \right|_{\epsilon^n}
 }
{\prod_j \left[ \left( l+q_i\right)^2 -\left( \mu^2+ m_i^2 \right)
  \right]},
\nonumber\\
\end{eqnarray}
In what follows we perform a separate reduction for each one of the CDR $\epsilon$
coefficients in the numerator. 
The dependence on $\epsilon$ due to the 
loop integration can only be made manifest if the integration over 
$\mu^2$ is performed. 
However, following the spirit of \cite{Anastasiou:2006jv}, we will avoid this integration. 

We now focus in the four dimensional inner integration. For every 
one-loop amplitude there exists a reduction 
\begin{eqnarray}
\label{eq:existsred}
\int d^4l
\frac{
\left. tr \left[ \Gamma_i^\dagger {\cal A}(l, \mu) \right]_{CDR} \right|_{\epsilon^n}
}
{\prod_j \left[ \left( l+q_i\right)^2 -\left( \mu^2+ m_i^2 \right) \right]} 
= \sum_k \Omega_k(\mu^2) {\rm Master}_k^{D=4}\left(\mu^2 \right),  
\end{eqnarray}
in terms of master integrals ${\rm Master}_k$ in the same dimension $D=4$. 
The master integrals in four dimensions are 
the scalar tadpole, bubble, triangle, and box integrals with unit numerator in their 
integrand. One can enhance however, the basis of master integrals 
by including the scalar pentagon. With this enhanced basis the coefficients of the 
master integrals are simple terminating polynomials in $\mu^2$ \cite{Anastasiou:2006jv,Britto:2008vq}: 
\begin{equation}
\label{eq:systemmu}
\Omega_k(\mu^2) = \Omega_k^{(0)} +  \mu^2 \Omega_k^{(1)} 
 +\ldots + (\mu^2)^ {{\rm max}} \Omega_k^{({\rm max})}. 
\end{equation} 
The  power $\bf {max}$ can be determined on simple dimensional grounds for 
each process and theory. We shall comment later on the methods that may be used to 
achieve this reduction. What is important to note now is that it can
be performed numerically,  since no divergence emerges.  
We then choose $\bf {max}$ different values of $\mu^2$ at each 
phase-space point and form a system of  equations from
Eq.~(\ref{eq:systemmu})\footnote{A similar inversion technique was previously used for a different purpose in Ref.~ \cite{Giele:2008ve}}. 
We solve the system numerically, and obtain the values  of the 
coefficients $\Omega_k^{(i)} $. 

With the $\Omega_k^{(i)} $ already evaluated it is easy to obtain the reduction 
in $D=4-2\epsilon$ dimensions. We write, 
\begin{eqnarray}
\label{xxxyyy}
&& 
\int d^Dl
\frac{
\left. tr \left[ \Gamma_i^\dagger {\cal A}(l, \mu) \right]_{CDR} \right|_{\epsilon^n}
}
{\prod_j \left[ \left( l+q_i\right)^2 -\left( \mu^2+ m_j^2 \right) \right]} 
\nonumber \\
&=&
\frac{-i}{\pi^2\Gamma(-\epsilon)}\int d\mu^2 
\left(\mu^2\right)^{-\epsilon-1} 
\int d^4l
\frac{
\left. tr \left[ \Gamma_i^\dagger {\cal A}(l, \mu) \right]_{CDR} \right|_{\epsilon^n}
}
{\prod_j \left[ \left( l+q_i\right)^2 -\left( \mu^2+ m_j^2 \right) \right]}  
\nonumber \\ 
&=& \sum_{k,i} \Omega_k^{(i)} \times 
\frac{-i}{\pi^2\Gamma(-\epsilon)}\int d\mu^2 
\left(\mu^2\right)^{-\epsilon-1} {\rm Master}_k^{D=4}\left(\mu^2 \right)  
\nonumber \\
&=& 
\sum_{k,i} \Omega_k^{(i)} \times {\rm Master}_k^{D=4-2\epsilon+2 i }\left(\mu^2=0 \right). 
\end{eqnarray}
In the last line we have achieved a reduction in terms of master integrals 
in $4-2\epsilon+2 i$ dimensions, where neither the master integrals nor the coefficients depend on the mass parameter
$\mu^2$.  These dimensionally shifted master integrals can be computed easily in terms of master integrals in $D=4-2\epsilon$ using the procedure described 
in Eq.~(\ref{eq:dimshift}). 

We now comment on the methods which we have employed to perform  
the four dimensional reduction of Eq.~(\ref{eq:existsred}). Our
application is simple enough so that an analytical or numerical reduction 
 using the program AIR~\cite{Anastasiou:2004vj} was possible.  
It is worth noting that the four
dimensional system of integration by parts identities which we need to
solve in Eq.~(\ref{eq:systemmu}) is much simpler than what it is
required for an equivalent direct reduction in D-dimensions.  
The two reductions involve the same number of symbolic parameters but they differ in complexity (
we have introduced a new mass parameter $\mu^2$ in
Eq.~(\ref{eq:existsred}) but at the same time we take $\epsilon =0$). 
This is mainly due to the fact that the reduction coefficients are a
simple polynomial in $\mu^2$ in the four dimensional reduction, 
while  in the D-dimensional reduction the coefficients are in general
rational functions. 

The best reduction method which we have used in Eq.~(\ref{eq:existsred}) 
is the one introduced by Ossola, Papadopoulos and 
Pittau~\cite{Ossola:2006us}. This method finds the 
coefficients of master integrals by evaluating the loop integrand at 
a finite number values of the loop momentum.  This is sufficient, since 
the integrand in four dimensions has a  known functional form 
as  a sum of  a small number of rational functions in
the loop momentum~\cite{Ossola:2006us}.

A different method than ours~\cite{Giele:2008ve} for a $D$-dimensional reduction also exploits the known functional form of one-loop
integrands~\cite{Ossola:2006us} and it computes the
coefficients of the  master integrals purely from tree amplitudes~\cite{Ellis:2007br}. Recently, the one-loop
amplitude for $gg \to t \bar{t} g$ was computed with this alternative 
method~\cite{Ellis:2008ir}. However, this method was used only in the FDH and the 'tHooft-Veltman schemes which we have decided to cautiously avoid as a first step. Finally, we note that our method is very close to what was  described
in~\cite{Anastasiou:2006jv}, although an analytic approach was envisaged there. In~\cite{Anastasiou:2006jv}, it was suggested that spinor
integration~\cite{Britto:2006sj} could also be used instead of the method of
~\cite{Ossola:2006us}  for the four dimensional part of the reduction.
The spinor integration method has developed significantly in the last
couple of years~\cite{Feng:2008ju,Britto:2008sw,Britto:2008vq,Britto:2007tt,Anastasiou:2006gt}. 
However, it remains an outstanding problem to compute the coefficient
of the tadpole master integral which vanishes when two propagators are
cut. 

\section{Results}
\label{sec:results}

As we discussed in previous sections, we calculated the NLO $gg\to Q\bar Q$ amplitude both analytically and fully numerically. In this section we are going to present both results for the unrenormalized amplitude. 

For our analytic results, we provide a \texttt{Mathematica} file, {\it mastercoeff.m}. We present the coefficients for the master integrals defined below for the unrenormalized amplitude with $N_l=0$, $N_h=0$ and $N=3$. The master integrals that appear in the amplitude are defined as
\allowdisplaybreaks[2]
\begin{align}
I_1 &= {\rm Tadp}(m_Q^2) =
\int\frac{d^Dk}{i\pi^{D/2}}\,\frac{1}{k^2-m_Q^2}\nonumber\\
I_2 &= {\rm Bub1}(s) =
\int\frac{d^Dk}{i\pi^{D/2}}\,\frac{1}{k^2}\,\frac{1}{(k+p_1+p_2)^2}\nonumber\\
I_3 &= {\rm Bub2}(m_Q^2,t) =
\int\frac{d^Dk}{i\pi^{D/2}}\,\frac{1}{k^2}\,\frac{1}{(k+p_2+p_3)^2-m_Q^2}\nonumber\\
I_4 &= {\rm Bub2}(m_Q^2,u) =
\int\frac{d^Dk}{i\pi^{D/2}}\,\frac{1}{k^2}\,\frac{1}{(k+p_1+p_3)^2-m_Q^2}\nonumber\\
I_5 &= {\rm Bub3}(m_Q^2,s) =
\int\frac{d^Dk}{i\pi^{D/2}}\,\frac{1}{k^2-m_Q^2}\,\frac{1}{(k+p_1+p_2)^2-m_Q^2}\nonumber\\
I_6 &= {\rm Tria1}(m_Q^2,s) =
\int\frac{d^Dk}{i\pi^{D/2}}\,\frac{1}{k^2}\,\frac{1}{(k+p_1+p_2)^2}\,\frac{1}{(k+p_1+p_2+p_3)^2-m_Q^2}\nonumber\\
I_7 &= {\rm Tria2}(m_Q^2,t) =
\int\frac{d^Dk}{i\pi^{D/2}}\,\frac{1}{k^2}\,\frac{1}{(k+p_3)^2-m_Q^2}\,\frac{1}{(k+p_2+p_3)^2-m_Q^2}\nonumber\\
I_8 &= {\rm Tria2}(m_Q^2,u) =
\int\frac{d^Dk}{i\pi^{D/2}}\,\frac{1}{k^2}\,\frac{1}{(k+p_3)^2-m_Q^2}\,\frac{1}{(k+p_1+p_3)^2-m_Q^2}\nonumber\\
I_9 &= {\rm Tria3}(m_Q^2,s) =
\int\frac{d^Dk}{i\pi^{D/2}}\,\frac{1}{k^2-m_Q^2}\,\frac{1}{(k+p_1)^2-m_Q^2}\,\frac{1}{(k+p_1+p_2)^2-m_Q^2}\nonumber\\
I_{10} &= {\rm Box1}(m_Q^2,s,t) =
\int\frac{d^Dk}{i\pi^{D/2}}\,\frac{1}{k^2}\frac{1}{(k+p_1)^2}\frac{1}{(k+p_1+p_2)^2}\frac{1}{(k+p_1+p_2+p_3)^2-m_Q^2}\nonumber\\
I_{11} &= {\rm Box1}(m_Q^2,s,u) =
\int\frac{d^Dk}{i\pi^{D/2}}\,\frac{1}{k^2}\frac{1}{(k+p_1)^2}\frac{1}{(k+p_1+p_2)^2}\frac{1}{(k-p_3)^2-m_Q^2}\nonumber\\
I_{12} &= {\rm Box2}(m_Q^2,s,t) =
\int\frac{d^Dk}{i\pi^{D/2}}\,\frac{1}{k^2}\frac{1}{(k+p_1)^2}\frac{1}{(k+p_1+p_3)^2-m_Q^2}\frac{1}{(k+p_1+p_2+p_3)^2-m_Q^2}\nonumber\\
I_{13} &= {\rm Box2}(m_Q^2,s,u) = \int\frac{d^Dk}{i\pi^{D/2}}\,\frac{1}{k^2}\frac{1}{(k+p_2)^2}\frac{1}{(k+p_2+p_3)^2-m_Q^2}\frac{1}{(k+p_1+p_2+p_3)^2-m_Q^2}\nonumber\\
I_{14} &= {\rm Box3}(m_Q^2,s,t) =
\int\frac{d^Dk}{i\pi^{D/2}}\,\frac{1}{k^2-m_Q^2}\frac{1}{(k+p_1)^2-m_Q^2}\frac{1}{(k+p_1+p_2)^2-m_Q^2}\frac{1}{(k+p_1+p_2+p_3)^2}\nonumber\\
I_{15} &= {\rm Box3}(m_Q^2,s,u) =
\int\frac{d^Dk}{i\pi^{D/2}}\,\frac{1}{k^2-m_Q^2}\frac{1}{(k+p_1)^2-m_Q^2}\frac{1}{(k+p_1+p_2)^2-m_Q^2}\frac{1}{(k-p_3)^2}\nonumber\\
\label{masters}
\end{align}
We used the results in Ref.~\cite{Korner:2004im} for numerical values of the master integrals defined above. 

In order to define the coefficients $B(i,L,j,k)$ in the file we provide, we write
\begin{equation}
d^{(1)}_{i,L} = \sum_{j=1}^{15}\,\sum_{k=-1}^4\,B(i,L,j,k)\,I_j\,\ep^k\,,
\end{equation}
where $d^{(1)}_{i,L}$ are defined in Eq.~(\ref{amplitude_sspace}), $I_j$ are the master integrals defined above and $k$ is the power of the dimensional regulator $\ep$. We give only the $B(i,L,j,k)$ needed for the $\mathcal{O}(\ep^2)$ expansion of the amplitude.

\begin{figure}[h]
\centerline{\epsfxsize=9cm \epsffile{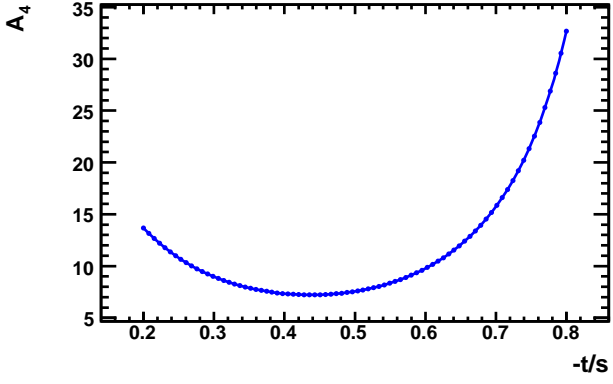}
\hspace{10mm} \epsfxsize=9cm \epsffile{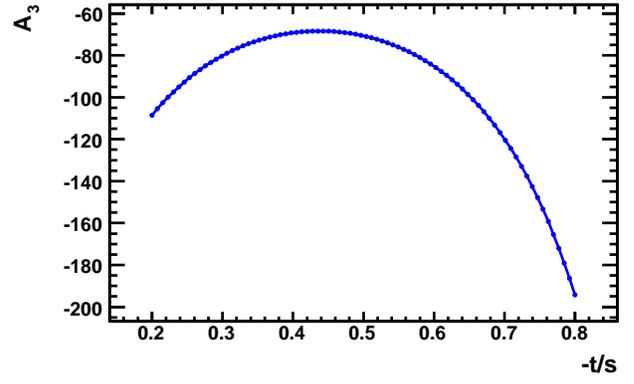}} \ \\
\hbox{\hskip 25mm (a) \hskip 96 mm (b)} \ \\
\centerline{\epsfxsize=9cm \epsffile{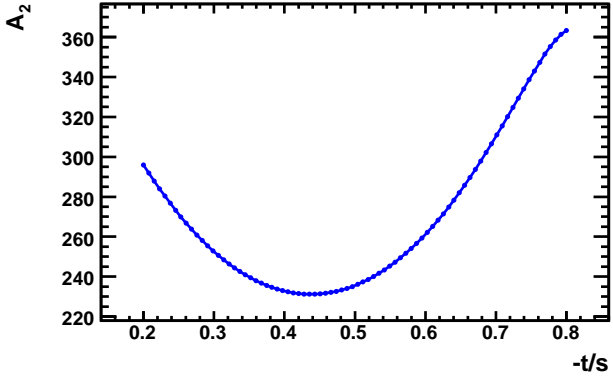}
\hspace{10mm} \epsfxsize=9cm \epsffile{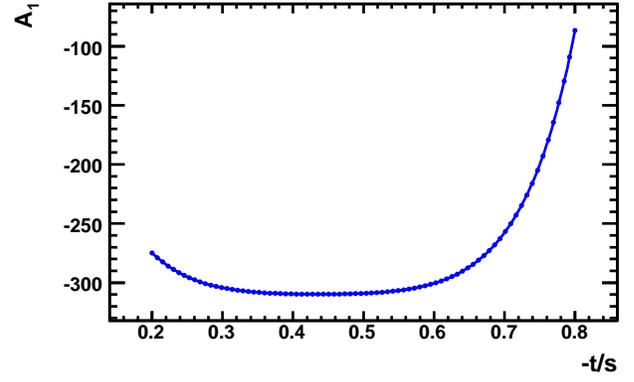}} \ \\
\hbox{\hskip 25mm (c) \hskip 96 mm (d)} \ \\
\centerline{\epsfxsize=9cm \epsffile{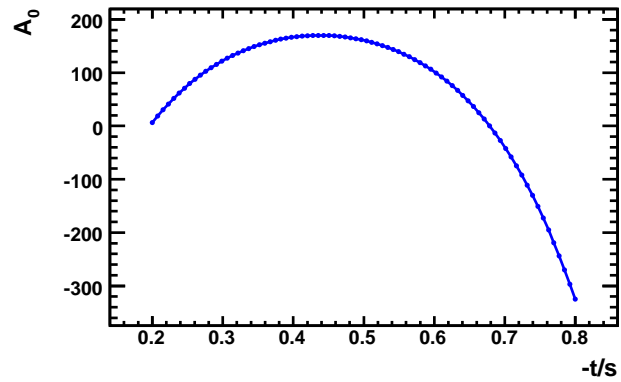} } \ \\
\hbox{\hskip 75mm (e)} \ \\
\caption{Results for (a) $A_4$ vs $-t/s$, (b) $A_3$ vs $-t/s$, (c) $A_2$ vs $-t/s$, (d) $A_1$ vs $-t/s$ and (e) $A_0$ vs $-t/s$. $A_4$, $A_3$, $A_2$, $A_1$ and $A_0$ are defined in Eq.~(\ref{A}).
\label{fig:results}}
\end{figure}

Our numerical results are presented in Fig.~\ref{fig:results}, where the
coefficients $A_i$  in Eq.~(\ref{A}) are plotted as a function of $-t/s$. 
In these plots, we have normalized  $m_Q=1$ and we chose
$s=16\,m_Q^2$. These numerical results agree of course with our
independent analytic evaluation. 
We have checked that the infrared poles of our amplitude agree  with
the  universal infrared pole structure  as predicted  in
Ref.~\cite{Catani:2000ef}. The finite part of the one-loop amplitude
agrees with the result of \cite{Korner:2002hy}. 

\begin{figure}[h!]
\centerline{\epsfxsize=5.5cm \epsffile{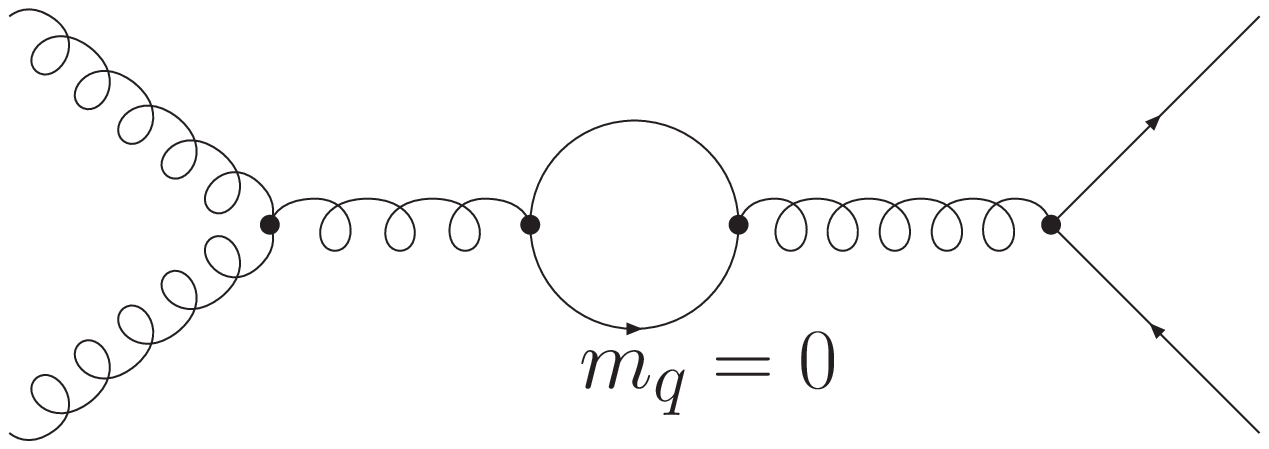}
\hspace{5mm} \epsfxsize=5.5cm \epsffile{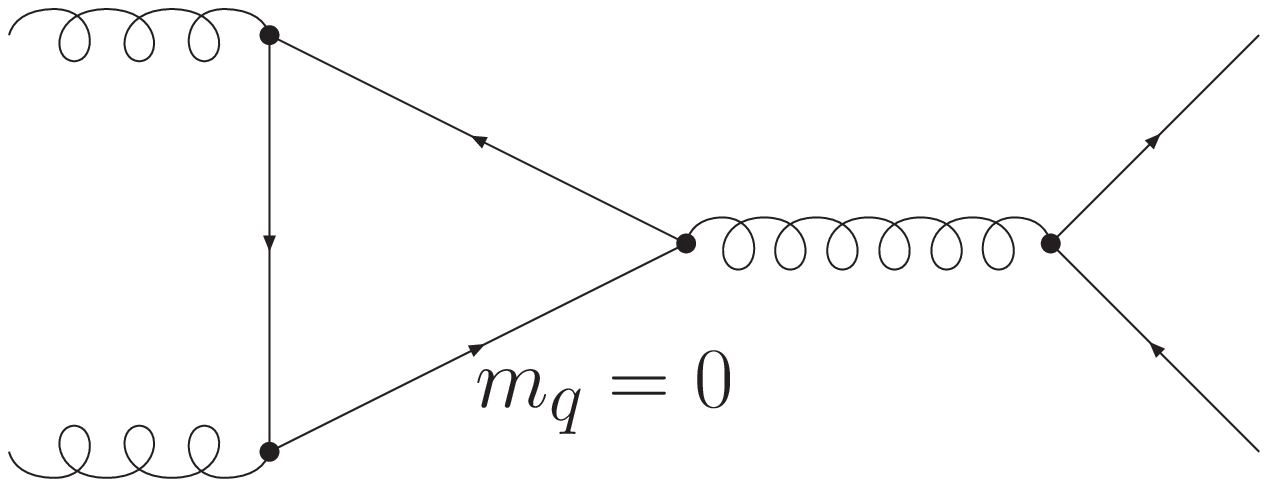}
\hspace{5mm} \epsfxsize=5.5cm \epsffile{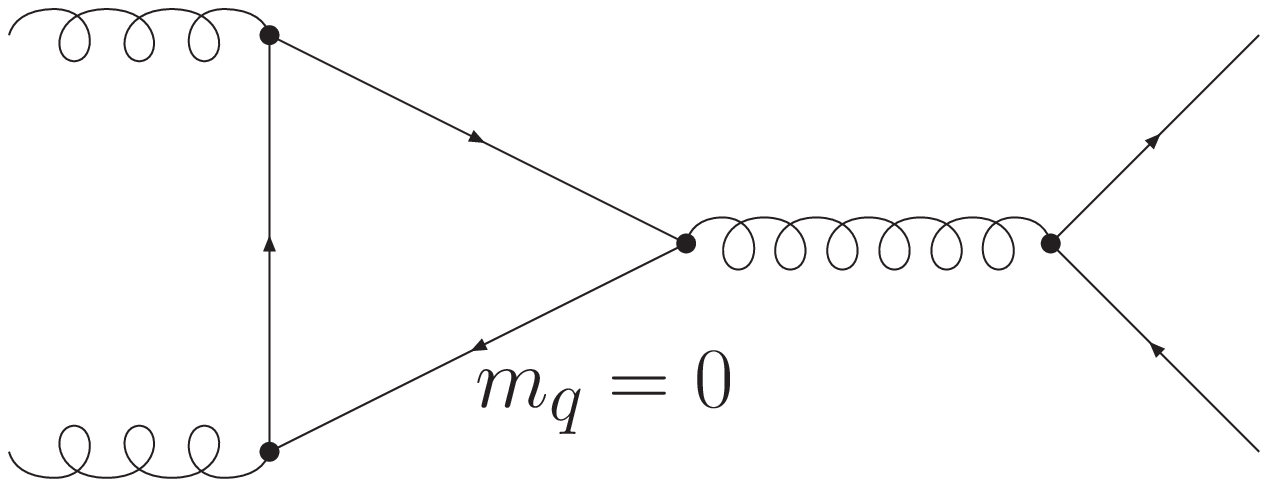}} \ \\
\hbox{\hskip 11mm (a) \hskip 61 mm (b) \hskip 57mm (c)} \ \\
\centerline{\epsfxsize=5.5cm \epsffile{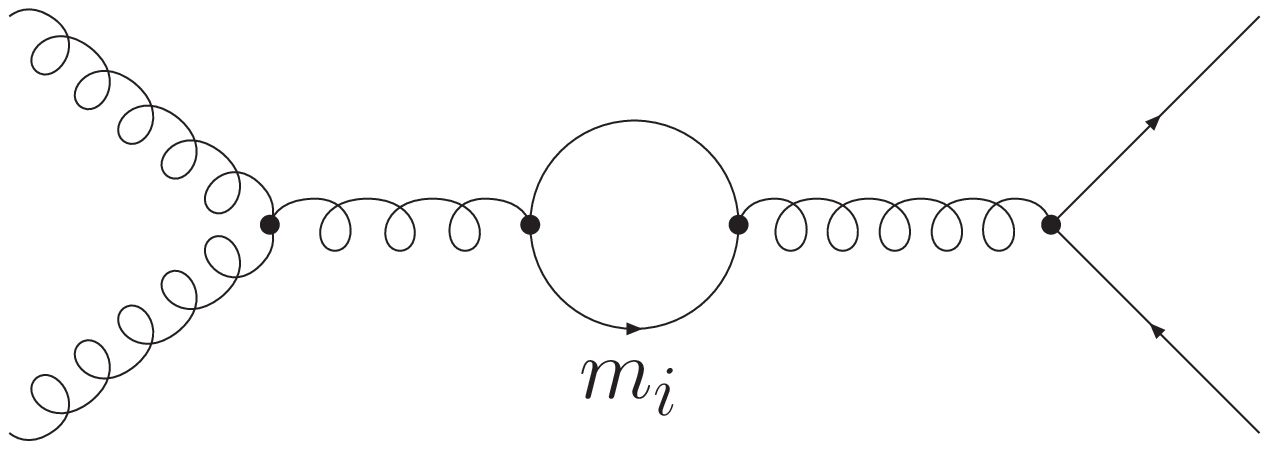}
\hspace{5mm} \epsfxsize=5.5cm \epsffile{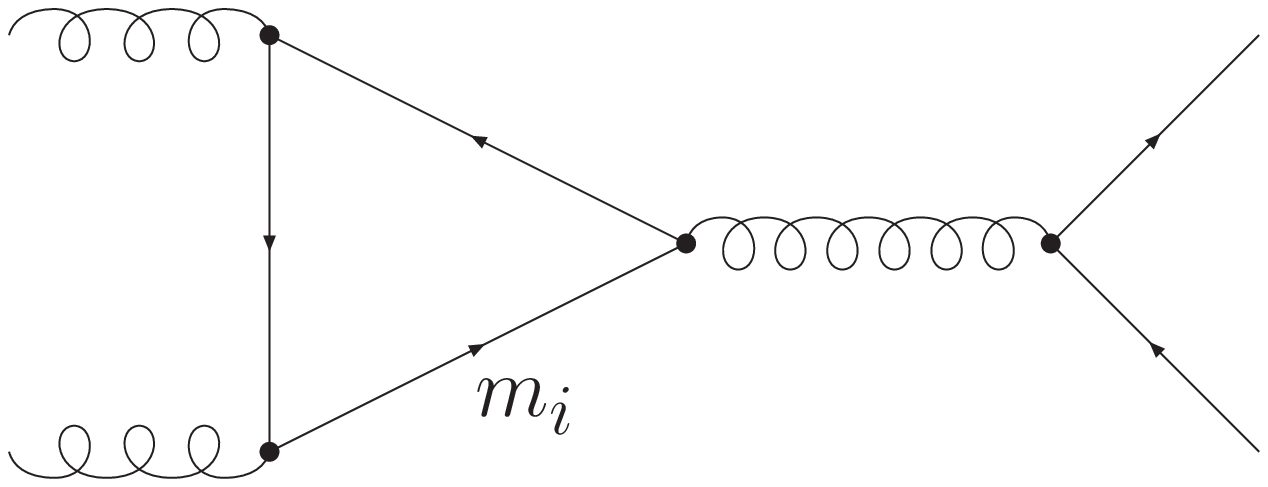}
\hspace{5mm} \epsfxsize=5.5cm \epsffile{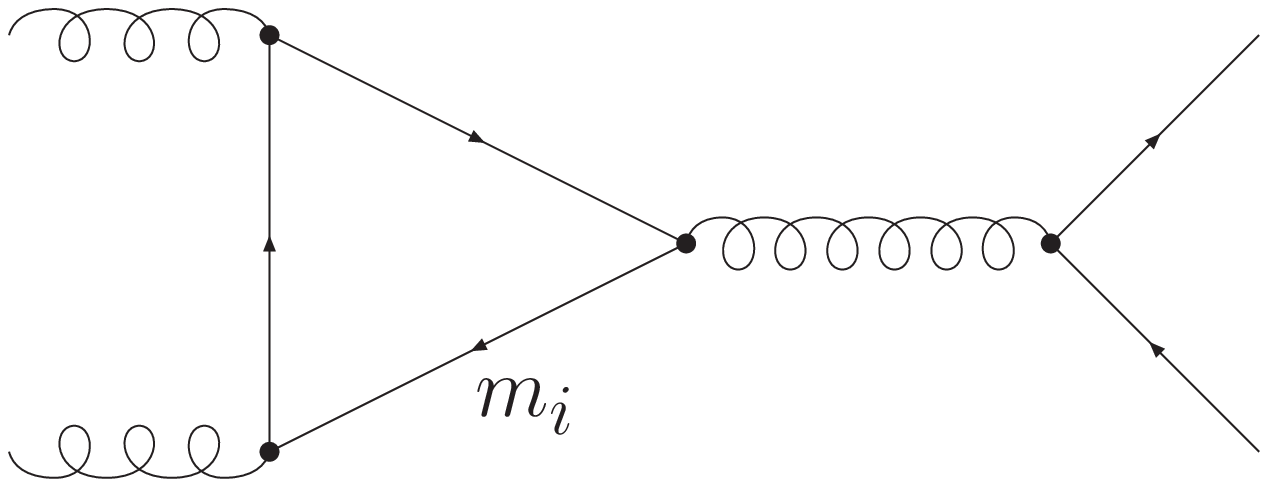}} \ \\
\hbox{\hskip 11mm (d) \hskip 61 mm (e) \hskip 57mm (f)} \ \\
\caption{(a)-(c) : Extra diagrams with internal massless quark loops. (d)-(f)
  : Extra diagrams with internal quark loops for quarks with mass $m_i$.
\label{fig:extra_diagrams}}
\end{figure}

The results we present in Fig.~\ref{fig:results} and in the
\texttt{Mathematica} file   correspond to zero number $N_l$ of light quarks
and zero number $N_h$ of heavy quarks with a different mass than the
external quark.  Contributions from  the diagrams shown in
Fig.~\ref{fig:extra_diagrams} need to be added separately. 
These extra terms read:
\begin{eqnarray}
M_{gg\to Q\bar Q}^{N_l} &=& \alpha_s^2\,\left(\frac{4\pi\mu^2}{m_Q^2}\right)^{\ep}\,N_l\,\, M_{gg\to Q\bar Q}^{(1),N_l}\,+\mathcal{O}(\alpha_s^3)\nonumber\\
M_{gg\to Q\bar Q}^{N_h} &=& \alpha_s^2\,\left(\frac{4\pi\mu^2}{m_Q^2}\right)^{\ep}\,\sum_{i=1}^{N_h}\,M_{gg\to Q\bar Q,i}^{(1)}\,+\mathcal{O}(\alpha_s^3)\,,
\end{eqnarray}
where 
\begin{eqnarray}
i\,M_{gg\to Q\bar Q}^{(1),N_l} &=& (m_Q^2)^{\ep}\,(c_1-c_2)\,\frac{2}{s}\,\frac{\ep}{(\ep-1)(2\ep-3)}\,{\rm Bub1}(s)\,\Gamma_3\nonumber\\
i\, M_{gg\to Q\bar Q,i}^{(1)} &=& (c_1-c_2)\frac{2}{s^2}\,\frac{1}{(\ep-1)(2\ep-3)}\,\Bigg(12\,{\rm Tadp}\,(m_i^2)\,(1-\ep)^2+2\,m_i^2\,s\,{\rm Tria3}\,(m_i^2,s)(2\ep-3)\nonumber\\
&\ &\hspace{5mm} -12\,m_i^2\,{\rm Bub3}\,(m_i^2,s)(1-\ep)-s\,{\rm Bub3}\,(m_i^2,s)\ep\Bigg)\,\Gamma_3\,.
\end{eqnarray}
where the definitions for the master integrals are given in
Eq.~(\ref{masters}). Note that in the above equations, $m_i$ are the masses of
the quarks for the internal loops and $m_Q$ is the mass of the external heavy
quarks. 

\section{Conclusions}
\label{sec:conclusions}
In this paper we compute the one-loop amplitude 
for $gg \to Q\bar{Q}$ through order ${\cal}(\epsilon^2)$ 
and its square through order ${\cal}(\epsilon^0)$. 
This is a contribution, albeit an easy one to compute, 
to the NNLO cross-section.    
We performed the computation in two different ways. We have 
developed  here a new reduction method for one-loop amplitudes 
which yields the full dependence of master integral coefficients in
the dimensional parameter in conventional dimensional regularization.
We are currently applying this method for computing other one-loop 
contributions at NNLO from processes with five external legs.

\clearpage

\begin{acknowledgments}
We would like to thank Zoltan Kunszt and Achilleas Lazopoulos for useful
comments. This work was supported by the Swiss National Science Foundation under contract 200021-117873.
\end{acknowledgments}

\begin{appendix}
\section{$\Gamma_{ij}$ and $\Gamma^{-1}_{ij}$}
\label{matrices}
In this appendix, we are going to give the elements of the matrices $\Gamma_{ij}$, defined in Eq.~(\ref{gamma}) and its inverse $\Gamma^{-1}_{ij}$. Note that both matrices are symmetric.

\allowdisplaybreaks[2]
\begin{align}
\Gamma_{11} &= 2 (-1 + \ep) (s + u + t) (s + 2 u + 2 t)\nonumber\\
\Gamma_{12} &= -\frac{1}{4 s}(s + u + t) (s + 2 u + 2 t)\nonumber\\
&\ \hspace{10mm}\times
 (s^2 + 2 s u +
     u^2 + 2 s t - 2 u t + t^2)\nonumber\\
\Gamma_{13} &= 2 (-1 + \ep) (u - t) (s + u + t)\nonumber\\
\Gamma_{14} &= -\frac{1}{4 s}(u - t) (s + u + t) (s^2 + 2 s u + u^2 + 
    2 s t - 2 u t + t^2)\nonumber\\
\Gamma_{15} &= -\frac{1}{s}(s + u + t) (s^2 + 2 s u + u^2 + 2 s t - 
    2 u t + t^2)\nonumber\\
\Gamma_{16} &= -\frac{1}{s}(s + u + t) (s^2 + 2 s u + u^2 + 2 s t - 
    2 u t + t^2)\nonumber\\
\Gamma_{17} &= \frac{1}{4} (s + u + t) (s^2 + 2 s u + u^2 + 2 s t - 
   2 u t + t^2)\nonumber\\
\Gamma_{18} &= \frac{1}{4} (s + u + t) (s^2 + 2 s u + u^2 + 2 s t - 
   2 u t + t^2)\nonumber\\
\Gamma_{19} &= 2 (-1 + \ep) (s + u + t) (s + 2 u + 2 t)\nonumber\\
\Gamma_{1\,10} &= -2 (-1 + \ep) (u - t) (s + u + t)\nonumber\\
\Gamma_{22} &=  -\frac{1}{16 s^2}(s + u + t) (s + 2 u + 2 t)\nonumber\\
&\ \hspace{10mm}\times (s^2 + 
    2 s u + u^2 + 2 s t - 2 u t + t^2)^2\nonumber\\
\Gamma_{23} &= -\frac{1}{4 s}(u - t) (s + u + t)\nonumber\\
&\ \hspace{10mm}\times (s^2 + 2 s u + u^2 + 
    2 s t - 2 u t + t^2)\nonumber\\
\Gamma_{24} &= -\frac{1}{16 s^2}(u - t) (s + u + t)\nonumber\\
&\ \hspace{10mm}\times (s^2 + 2 s u + 
    u^2 + 2 s t - 2 u t + t^2)^2\nonumber\\
\Gamma_{25} &= -\frac{1}{4 s^2}(s + u + t) (s^2 + 2 s u + u^2 + 
    2 s t - 2 u t + t^2)^2\nonumber\\
\Gamma_{26} &= -\frac{1}{4 s^2}(s + u + t) (s^2 + 2 s u + u^2 + 
    2 s t - 2 u t + t^2)^2\nonumber\\
\Gamma_{27} &= \frac{1}{16 s}(s + u + t) (s^2 + 2 s u + u^2 + 2 s t - 
   2 u t + t^2)^2\nonumber\\
\Gamma_{28} &= \frac{1}{16 s}(s + u + t) (s^2 + 2 s u + u^2 + 2 s t - 
   2 u t + t^2)^2\nonumber\\
\Gamma_{29} &= -\frac{1}{4 s}(s + u + t) (s + 2 u + 2 t)\nonumber\\
&\ \hspace{10mm}\times (s^2 + 2 s u +
     u^2 + 2 s t - 2 u t + t^2)\nonumber\\
\Gamma_{2\,10} &= \frac{1}{4 s}(u - t) (s + u + t) (s^2 + 2 s u + u^2 + 
   2 s t - 2 u t + t^2)\nonumber\\
\Gamma_{33} &= -(-1 + \ep) (s + u - t) (s - u + t)\nonumber\\
\Gamma_{34} &= \frac{1}{8 s}(s + u - t) (s - u + t)\nonumber\\
&\ \hspace{10mm}\times (s^2 + 2 s u + 
   u^2 + 2 s t - 2 u t + t^2)\nonumber\\
\Gamma_{35} &= -\frac{1}{2s}((u - t) (s^2 + 2 s u + u^2 + 2 s t - 2 u t + 
    t^2))\nonumber\\
\Gamma_{36} &= -\frac{1}{2s}((u - t) (s^2 + 2 s u + u^2 + 2 s t - 2 u t + 
    t^2))\nonumber\\
\Gamma_{37} &= 0\nonumber\\
\Gamma_{38} &= 0\nonumber\\
\Gamma_{39} &= 2 (-1 + \ep) (u - t) (s + u + t)\nonumber\\
\Gamma_{3\,10} &= (-1 + \ep) (s + u - t) (s - u + t)\nonumber\\
\Gamma_{44} &= \frac{1}{32 s^2}(s + u - t) (s - u +
t)\nonumber\\
&\ \hspace{10mm}\times (s^2 + 2 s u +
    u^2 + 2 s t - 2 u t + t^2)^2\nonumber\\
\Gamma_{45} &= -\frac{1}{8s^2}((u - t) (s^2 + 2 s u + u^2 + 2 s t - 2 u t + 
    t^2)^2)\nonumber\\
\Gamma_{46} &= -\frac{1}{8s^2}((u - t) (s^2 + 2 s u + u^2 + 2 s t - 2 u t + 
    t^2)^2)\nonumber\\
\Gamma_{47} &= 0\nonumber\\
\Gamma_{48} &= 0\nonumber\\
\Gamma_{49} &= -\frac{1}{4s}((u - t) (s + u + t) (s^2 + 2 s u + u^2 + 
    2 s t - 2 u t + t^2))\nonumber\\
\Gamma_{4\,10} &= -\frac{1}{8 s} (s + u - t) (s - u +
t)\nonumber\\ 
&\ \hspace{10mm}\times (s^2 + 2 s u + 
    u^2 + 2 s t - 2 u t + t^2)\nonumber\\
\Gamma_{55} &= \frac{1}{2 s^2}(-3 s^2 + 2 \ep s^2 - 2 s u - u^2 - 
   2 s t + 2 u t - t^2)\nonumber\\ 
&\ \hspace{10mm}\times (s^2 + 2 s u + u^2 + 
   2 s t - 2 u t + t^2)\nonumber\\
\Gamma_{56} &= -\frac{1}{2 s^2}(s^2 + 2 s u + u^2 + 2 s t - 2 u t + 
    t^2)\nonumber\\ 
&\ \hspace{10mm}\times (2 s^2 + 2 s u + u^2 + 2 s t - 2 u t + 
    t^2)\nonumber\\
\Gamma_{57} &= -\frac{1}{2} (-1 + \ep) (s + u + t) (s^2 + 2 s u + u^2 + 
   2 s t - 2 u t + t^2)\nonumber\\
\Gamma_{58} &= \frac{1}{4} (s + u + t) (s^2 + 2 s u + u^2 + 2 s t - 
   2 u t + t^2)\nonumber\\
\Gamma_{59} &= -\frac{1}{s}((s + u + t) (s^2 + 2 s u + u^2 + 2 s t - 
    2 u t + t^2))\nonumber\\
\Gamma_{5\,10} &= \frac{1}{2s}((s - 2 \ep s + u - t) (s^2 + 2 s u + u^2 + 
   2 s t - 2 u t + t^2))\nonumber\\
\Gamma_{66} &= \frac{1}{2 s^2}(-3 s^2 + 2 \ep s^2 - 2 s u - u^2 - 
   2 s t + 2 u t - t^2)\nonumber\\ 
&\ \hspace{10mm}\times (s^2 + 2 s u + u^2 + 
   2 s t - 2 u t + t^2)\nonumber\\
\Gamma_{67} &= \frac{1}{4} (s + u + t) (s^2 + 2 s u + u^2 + 2 s t - 
   2 u t + t^2)\nonumber\\
\Gamma_{68} &= -\frac{1}{2} (-1 + \ep) (s + u + t) (s^2 + 2 s u + u^2 + 
   2 s t - 2 u t + t^2)\nonumber\\
\Gamma_{69} &= -\frac{1}{s}((s + u + t) (s^2 + 2 s u + u^2 + 2 s t - 
    2 u t + t^2))\nonumber\\
\Gamma_{6\,10} &= \frac{1}{2s}((-s + 2 \ep s + u - t) (s^2 + 2 s u + u^2 + 
   2 s t - 2 u t + t^2))\nonumber\\
\Gamma_{77} &= \frac{1}{8 s}(-1 + \ep) (s + u - t) (s - u +
t)\nonumber\\ 
&\ \hspace{10mm}\times (s + u +
    t) (s^2 + 2 s u + u^2 + 2 s t - 2 u t +
    t^2)\nonumber\\
\Gamma_{78} &= -\frac{1}{16 s}(s + u - t) (s - u +
t)\nonumber\\ 
&\ \hspace{10mm}\times (s + u + 
    t) (s^2 + 2 s u + u^2 + 2 s t - 2 u t +
    t^2)\nonumber\\
\Gamma_{79} &= -\frac{1}{4 s}(s + u + t) (-s - u + 2 \ep u + t - 
    2 \ep t)\nonumber\\ 
&\ \hspace{10mm}\times (s^2 + 2 s u + u^2 + 2 s t - 2 u t + 
    t^2)\nonumber\\
\Gamma_{7\,10} &= 0\nonumber\\
\Gamma_{88} &= \frac{1}{8 s}(-1 + \ep) (s + u - t) (s - u +
t)\nonumber\\ 
&\ \hspace{10mm}\times (s + u +
    t) (s^2 + 2 s u + u^2 + 2 s t - 2 u t +
    t^2)\nonumber\\
\Gamma_{89} &=\frac{1}{4 s}(s + u + t) (s - u + 2 \ep u + t - 
   2 \ep t)\nonumber\\ 
&\ \hspace{10mm}\times (s^2 + 2 s u + u^2 + 2 s t - 2 u t + 
   t^2)\nonumber\\
\Gamma_{8\,10} &= 0\nonumber\\
\Gamma_{99} &= -\frac{1}{s} 2 (s + u + t)\nonumber\\
&\ \hspace{5mm}\times (s^2 - 2 \ep s^2 + 2 \ep^2 s^2 + 
    2 s u - 4 \ep s u + 4 \ep^2 s u - u^2\nonumber\\
&\ \hspace{10mm} + 
    2 \ep u^2 + 2 s t - 4 \ep s t + 4 \ep^2 s t + 
    2 u t - 4 \ep u t - t^2 + 2 \ep t^2)\nonumber\\
\Gamma_{9\,10} &= 4 (-1 + \ep)^2 (u - t) (s + u + t)\nonumber\\
\Gamma_{10\,10} &= 2 (-1 + \ep)^2 (s + u - t) (s - u + t)
\end{align}

\begin{align}
\Gamma^{-1}_{11} &= \frac{(s + u - t) (s - u + t)}{2 \ep s (s + u + 
     t) (s^2 + 2 s u + u^2 + 2 s t - 2 u t + t^2)}\nonumber\\
\Gamma^{-1}_{12} &= -\frac{4 (s + u - t) (s - u + t)}{(-1 + 2 \ep) (s + u + 
      t) (s^2 + 2 s u + u^2 + 2 s t - 2 u t + 
      t^2)^2}\nonumber\\
\Gamma^{-1}_{13} &= \frac{u - t}{\ep s (s^2 + 2 s u + u^2 + 2 s t - 
   2 u t + t^2)}\nonumber\\
\Gamma^{-1}_{14} &= -\frac{8 (u - t)}{(-1 + 2 \ep) (s^2 + 2 s u + u^2 + 
    2 s t - 2 u t + t^2)^2}\nonumber\\
\Gamma^{-1}_{15} &= 0\nonumber\\
\Gamma^{-1}_{16} &= 0\nonumber\\
\Gamma^{-1}_{17} &= \frac{2 (u - t)}{\ep (-1 + 2 \ep) s (s + u + t) (s^2 + 
     2 s u + u^2 + 2 s t - 2 u t + t^2)}\nonumber\\
\Gamma^{-1}_{18} &= -\frac{2 (u - t)}{\ep (-1 + 2 \ep) s (s + u + t) (s^2 + 
      2 s u + u^2 + 2 s t - 2 u t + t^2)}\nonumber\\
\Gamma^{-1}_{19} &= \frac{(s + u - t) (s - u + t)}{2 \ep (-1 + 2 \ep) s (s + 
     u + t) (s^2 + 2 s u + u^2 + 2 s t - 2 u t + 
     t^2)}\nonumber\\
\Gamma^{-1}_{1\,10} &= -\frac{u - t}{\ep (-1 + 2 \ep) s (s^2 + 2 s u + u^2 + 
      2 s t - 2 u t + t^2)}\nonumber\\
\Gamma^{-1}_{22} &= -\frac{32 (-2 + \ep) s (s + u - t) (s - u + t)}{(-1 + 
      2 \ep) (s + u + t) (s^2 + 2 s u + u^2 + 
      2 s t - 2 u t + t^2)^3}\nonumber\\
\Gamma^{-1}_{23} &= -\frac{8 (u - t)}{(-1 + 2 \ep) (s^2 + 2 s u + u^2 + 
    2 s t - 2 u t + t^2)^2}\nonumber\\
\Gamma^{-1}_{24} &= -\frac{64 (-2 + \ep) s (u - t)}{(-1 + 2 \ep) (s^2 + 2 s u + 
    u^2 + 2 s t - 2 u t + t^2)^3}\nonumber\\
\Gamma^{-1}_{25} &= 0\nonumber\\
\Gamma^{-1}_{26} &= 0\nonumber\\
\Gamma^{-1}_{27} &= \frac{16 s}{(-1 + 2 \ep) (s + u + t) (s^2 + 2 s u + u^2 + 
     2 s t - 2 u t + t^2)^2}\nonumber\\
\Gamma^{-1}_{28} &= \frac{16 s}{(-1 + 2 \ep) (s + u + t) (s^2 + 2 s u + u^2 + 
     2 s t - 2 u t + t^2)^2}\nonumber\\
\Gamma^{-1}_{29} &= 0\nonumber\\
\Gamma^{-1}_{2\,10} &= 0\nonumber\\
\Gamma^{-1}_{33} &= -\frac{s + 2 u + 2 t}{
 \ep s (s^2 + 2 s u + u^2 + 2 s t - 2 u t + t^2)}\nonumber\\
\Gamma^{-1}_{34} &= \frac{8 (s + 2 u + 2 t)}{(-1 + 2 \ep) (s^2 + 2 s u + u^2 + 
   2 s t - 2 u t + t^2)^2}\nonumber\\
\Gamma^{-1}_{35} &= -\frac{1}{\ep (-1 + 2 \ep) (s^2 + 2 s u + u^2 + 2 s t - 
      2 u t + t^2)}\nonumber\\
\Gamma^{-1}_{36} &= \frac{1}{\ep (-1 + 2 \ep) (s^2 + 2 s u + u^2 + 2 s t - 
     2 u t + t^2)}\nonumber\\
\Gamma^{-1}_{37} &= -\frac{4}{\ep (-1 + 2 \ep) s (s^2 + 2 s u + u^2 + 2 s t - 
      2 u t + t^2)}\nonumber\\
\Gamma^{-1}_{38} &= \frac{4}{\ep (-1 + 2 \ep) s (s^2 + 2 s u + u^2 + 2 s t - 
     2 u t + t^2)}\nonumber\\
\Gamma^{-1}_{39} &= \frac{u - t}{\ep (-1 + 2 \ep) s (s^2 + 2 s u + u^2 + 
     2 s t - 2 u t + t^2)}\nonumber\\
\Gamma^{-1}_{3\,10} &= \frac{s + 2 u + 
   2 t}{\ep (-1 + 2 \ep) s (s^2 + 2 s u + u^2 + 
     2 s t - 2 u t + t^2)}\nonumber\\
\Gamma^{-1}_{44} &= \frac{64 (-s^2 + \ep s^2 - 2 s u + 2 \ep s u + u^2 - 
     2 s t + 2 \ep s t - 2 u t + t^2)}{(-1 + 
     2 \ep) (s^2 + 2 s u + u^2 + 2 s t - 2 u t + 
     t^2)^3}\nonumber\\
\Gamma^{-1}_{45} &= -\frac{8 (u - t)}{(-1 + 2 \ep) (s^2 + 2 s u + u^2 + 
    2 s t - 2 u t + t^2)^2}\nonumber\\
\Gamma^{-1}_{46} &= -\frac{8 (u - t)}{(-1 + 2 \ep) (s^2 + 2 s u + u^2 + 
    2 s t - 2 u t + t^2)^2}\nonumber\\
\Gamma^{-1}_{47} &= 0\nonumber\\
\Gamma^{-1}_{48} &= 0\nonumber\\
\Gamma^{-1}_{49} &= 0\nonumber\\
\Gamma^{-1}_{4\,10} &= 0\nonumber\\
\Gamma^{-1}_{55} &= -\frac{(s + u - t) (s - u + t)}{
 \ep (s^2 + 2 s u + u^2 + 2 s t - 2 u t + t^2)^2}\nonumber\\
\Gamma^{-1}_{56} &= -\frac{(s + u - t) (s - u + t)}{\ep (-1 + 2 \ep) (s^2 + 
      2 s u + u^2 + 2 s t - 2 u t + t^2)^2}\nonumber\\
\Gamma^{-1}_{57} &= -\frac{4 s}{
 \ep (s^2 + 2 s u + u^2 + 2 s t - 2 u t + t^2)^2}\nonumber\\
\Gamma^{-1}_{58} &= -\frac{4 s}{\ep (-1 + 2 \ep) (s^2 + 2 s u + u^2 + 2 s t - 
      2 u t + t^2)^2}\nonumber\\
\Gamma^{-1}_{59} &= 0\nonumber\\
\Gamma^{-1}_{5\,10} &= -\frac{1}{\ep (-1 + 2 \ep) (s^2 + 2 s u + u^2 + 2 s t - 
      2 u t + t^2)}\nonumber\\
\Gamma^{-1}_{66} &= -\frac{(s + u - t) (s - u + t)}{
 \ep (s^2 + 2 s u + u^2 + 2 s t - 2 u t + t^2)^2}\nonumber\\
\Gamma^{-1}_{67} &= -\frac{4 s}{\ep (-1 + 2 \ep) (s^2 + 2 s u + u^2 + 2 s t - 
      2 u t + t^2)^2}\nonumber\\
\Gamma^{-1}_{68} &= -\frac{4 s}{
 \ep (s^2 + 2 s u + u^2 + 2 s t - 2 u t + t^2)^2}\nonumber\\
\Gamma^{-1}_{69} &= 0\nonumber\\
\Gamma^{-1}_{6\,10} &= \frac{1}{\ep (-1 + 2 \ep) (s^2 + 2 s u + u^2 + 2 s t - 
     2 u t + t^2)}\nonumber\\
\Gamma^{-1}_{77} &= -\frac{8 (-2 s^2 + 2 \ep s^2 - 2 s u - u^2 - 2 s t + 
      2 u t - t^2)}{\ep (-1 + 2 \ep) s (s + u + 
      t) (s^2 + 2 s u + u^2 + 2 s t - 2 u t + 
      t^2)^2}\nonumber\\
\Gamma^{-1}_{78} &= -\frac{8 (2 s^2 + 2 s u + u^2 + 2 s t - 2 u t + 
      t^2)}{\ep (-1 + 2 \ep) s (s + u + t) (s^2 + 
      2 s u + u^2 + 2 s t - 2 u t + t^2)^2}\nonumber\\
\Gamma^{-1}_{79} &= -\frac{2 (u - t)}{\ep (-1 + 2 \ep) s (s + u + t) (s^2 + 
      2 s u + u^2 + 2 s t - 2 u t + t^2)}\nonumber\\
\Gamma^{-1}_{7\,10} &= -\frac{4}{\ep (-1 + 2 \ep) s (s^2 + 2 s u + u^2 + 2 s t - 
      2 u t + t^2)}\nonumber\\
\Gamma^{-1}_{88} &= -\frac{8 (-2 s^2 + 2 \ep s^2 - 2 s u - u^2 - 2 s t + 
      2 u t - t^2)}{\ep (-1 + 2 \ep) s (s + u + 
      t) (s^2 + 2 s u + u^2 + 2 s t - 2 u t + 
      t^2)^2}\nonumber\\
\Gamma^{-1}_{89} &= \frac{2 (u - t)}{\ep (-1 + 2 \ep) s (s + u + t) (s^2 + 
     2 s u + u^2 + 2 s t - 2 u t + t^2)}\nonumber\\
\Gamma^{-1}_{8\,10} &= \frac{4}{\ep (-1 + 2 \ep) s (s^2 + 2 s u + u^2 + 2 s t - 
     2 u t + t^2)}\nonumber\\
\Gamma^{-1}_{99} &= -\frac{(s + u - t) (s - u + t)}{2 \ep (-1 + 2 \ep) s (s + 
      u + t) (s^2 + 2 s u + u^2 + 2 s t - 2 u t + 
      t^2)}\nonumber\\
\Gamma^{-1}_{9\,10} &= \frac{u - t}{\ep (-1 + 2 \ep) s (s^2 + 2 s u + u^2 + 
     2 s t - 2 u t + t^2)}\nonumber\\
\Gamma^{-1}_{10\,10} &= \frac{s + 2 u + 
   2 t}{\ep (-1 + 2 \ep) s (s^2 + 2 s u + u^2 + 
     2 s t - 2 u t + t^2)}\,.
\end{align}


\section{Generic Dimensional Shift Identities}\label{app_IBP}

In this appendix we are going to give generic dimensional shift identities for box, triangle and bubble topologies with arbitrary masses for different propagators. For this purpose let's define the following scalar integrals
\begin{eqnarray}
{\rm Box}(m_0,m_1,m_2,m_3,q_0,q_1,q_2,q_3;D)&\equiv& \int\frac{d^Dk}{i\pi^{D/2}}\frac{1}{[(k+q_0)^2-m_0^2]}\frac{1}{[(k+q_1)^2-m_1^2]}\nonumber\\
&\ &\hspace{40mm}\times \frac{1}{[(k+q_2)^2-m_2^2]}\frac{1}{[(k+q_3)^2-m_3^2]}\,,\nonumber\\\nonumber\\
{\rm Tri}(m_0,m_1,m_2,q_0,q_1,q_2;D)&\equiv& \int\frac{d^Dk}{i\pi^{D/2}}\frac{1}{[(k+q_0)^2-m_0^2]}\frac{1}{[(k+q_1)^2-m_1^2]}\frac{1}{[(k+q_2)^2-m_2^2]}\,,\nonumber\\\nonumber\\
{\rm Bub}(m_0,m_1,q_0,q_1;D)&\equiv& \int\frac{d^Dk}{i\pi^{D/2}}\frac{1}{[(k+q_0)^2-m_0^2]}\frac{1}{[(k+q_1)^2-m_1^2]}\,,\nonumber\\\nonumber\\
{\rm Tadp}(m;D)&\equiv&\int\frac{d^Dk}{i\pi^{D/2}}\frac{1}{k^2-m_0^2}\,,
\end{eqnarray}
where in the definitions of the scalar integrals, the order of the arguments correspond one to one to the order of the propagators.

For the bubble topology we have
\begin{eqnarray}
{\rm Bub}(m_0,m_1,0,q_1;D+2) &=& \frac{1}{2 (D-1) q_1^2}\nonumber\\
&\ &\hspace{-50mm}\times\Bigg\{{\rm Bub}(m_0,m_1,0,q_1;D)\Bigg[m_0^4 + (m_1^2 - q_1^2)^2 - 2 m_0^2 (m_1^2 + q_1^2)\Bigg]\nonumber\\
&\ &\hspace{-50mm} + 
  m_1^2 \Bigg[-{\rm Tadp}(m_1;D) + {\rm Tadp}(m_0;D)\Bigg] - m_0^2 \Bigg[-{\rm Tadp}(m_1;D) +  {\rm Tadp}(m_0;D)\Bigg]\nonumber\\
&\ &\hspace{-50mm} - 
  q_1^2 \Bigg[{\rm Tadp}(m_1;D) +  {\rm Tadp}(m_0;D)\Bigg]\Bigg\}\,.\nonumber\\
\end{eqnarray}
For the triangle topology
\begin{eqnarray}
{\rm Tri}(m_0,m_1,m_2,0,q_1,q_2;D+2) &=& \frac{1}{(D-2) \Big[q_1^4 + (q_2^2 - q_{12}^2)^2 - 2 q_1^2 (q_2^2 + q_{12}^2)\Big]}\nonumber\\
&\ &\hspace{-70mm}\times\Bigg\{{\rm Bub}(m_1,m_2,q_1,q_2;D)\Bigg[-(m_1^2 - m_2^2) (q_1^2 - q_2^2) + (-2 m_0^2 + m_1^2 + m_2^2 + 
       q_1^2 + q_2^2) q_{12}^2 - q_{12}^4\Bigg]\nonumber\\
&\ &\hspace{-70mm} + 
 {\rm Bub}(m_0,m_2,0,q_2;D)\Bigg[m_2^2 (q_1^2 + q_2^2 - q_{12}^2) + 
    q_2^2 (-2 m_1^2 + q_1^2 - q_2^2 + q_{12}^2) + 
    m_0^2 (-q_1^2 + q_2^2 + q_{12}^2)\Bigg]\nonumber\\
&\ &\hspace{-70mm} + 
 {\rm Bub}(m_0,m_1,0,q_1;D)\Bigg[m_1^2 (q_1^2 + q_2^2 - q_{12}^2) + m_0^2 (q_1^2 - q_2^2 + q_{12}^2) +     q_1^2 (-2 m_2^2 - q_1^2 + q_2^2 + q_{12}^2)\Bigg]\nonumber\\
&\ &\hspace{-70mm} - 
 2\, {\rm Tri}(m_0,m_1,m_2,0,q_1,q_2;D)\Bigg[m_1^4 q_2^2 + m_0^4 q_{12}^2\nonumber\\
&\ &\hspace{-50mm} + 
    m_0^2 \Big[(m_1^2 - m_2^2) (q_1^2 - q_2^2) - (m_1^2 + m_2^2 + q_1^2 + 
          q_2^2) q_{12}^2 + q_{12}^4\Big]\nonumber\\
&\ &\hspace{-50mm} + 
    q_1^2 \Big[m_2^4 + q_2^2 q_{12}^2 + m_2^2 (q_1^2 - q_2^2 - q_{12}^2)\Big] - 
    m_1^2 \Big[m_2^2 (q_1^2 + q_2^2 - q_{12}^2) + q_2^2 (q_1^2 - q_2^2 + q_{12}^2)\Big]\Bigg] \Bigg\}\,,\nonumber\\
\end{eqnarray}
where we defined
\begin{equation}
q_{ij}^2 = (q_j-q_i)^2\,.
\end{equation}
For the box topology, the most general expression is quite lengthy. Therefore we will give the answer for $q_1^2=0$
\allowdisplaybreaks[2]
\begin{align}
{\rm Box}(m_0,m_1,m_2,m_3,0,q_1,q_2,q_3;D+2) &=\nonumber\\\nonumber\\
&\ \hspace{-80mm} \frac{1}{2 (D-3)\Big[q_{12}^4 q_3^2 + 
   q_{12}^2 (q_{13}^2 (-q_2^2 + q_{23}^2) - (q_{13}^2 + q_2^2 + q_{23}^2) q_3^2 + 
      q_3^4) + q_2^2 (q_{13}^4 + q_{23}^2 q_3^2 + q_{13}^2 (q_2^2 - q_{23}^2 - q_3^2))\Big]}\nonumber\\
&\ \hspace{-80mm}\times\Bigg\{{\rm Tri}(m_0,m_1,m_3,0,q_1,q_3;D)\Bigg[-m_3^2 (q_{12}^2 - q_2^2) (q_{13}^2 - q_3^2) + 
    m_2^2 (q_{13}^2 - q_3^2)^2 - (q_{13}^2 - q_3^2) (q_{13}^2 q_2^2 - q_{12}^2 q_3^2)\nonumber\\
&\ \hspace{-70mm} +
     m_0^2 (q_{13}^2 (q_{12}^2 - q_{13}^2 - 2 q_2^2 + q_{23}^2) + (q_{12}^2 + q_{13}^2 - 
          q_{23}^2) q_3^2) + 
    m_1^2 (q_3^2 (-2 q_{12}^2 + q_2^2 + q_{23}^2 - q_3^2)\nonumber\\
&\ \hspace{-70mm} + 
       q_{13}^2 (q_2^2 - q_{23}^2 + q_3^2))\Bigg]\nonumber\\
&\ \hspace{-80mm} + 
 {\rm Tri}(m_1,m_2,m_3,q_1,q_2,q_3;D)\Bigg[q_{12}^2 q_{13}^2 q_2^2 - q_{13}^4 q_2^2 - 2 q_{12}^2 q_{13}^2 q_{23}^2 + 
    q_{13}^2 q_2^2 q_{23}^2\nonumber\\
&\ \hspace{-70mm} + 
    m_0^2 (q_{12}^4 + (q_{13}^2 - q_{23}^2)^2 - 2 q_{12}^2 (q_{13}^2 + q_{23}^2)) - 
    q_{12}^4 q_3^2 + q_{12}^2 q_{13}^2 q_3^2 + q_{12}^2 q_{23}^2 q_3^2\nonumber\\
&\ \hspace{-70mm} + 
    m_2^2 (q_{13}^2 (q_{12}^2 - q_{13}^2 - 2 q_2^2 + q_{23}^2) + (q_{12}^2 + q_{13}^2 - 
          q_{23}^2) q_3^2)\nonumber\\
&\ \hspace{-70mm} + 
    m_3^2 (-q_{12}^4 + q_2^2 (q_{13}^2 - q_{23}^2) + 
       q_{12}^2 (q_{13}^2 + q_2^2 + q_{23}^2 - 2 q_3^2))\nonumber\\
&\ \hspace{-70mm} + 
    m_1^2 (q_{13}^2 (q_2^2 + q_{23}^2 - q_3^2) + q_{23}^2 (q_2^2 - q_{23}^2 + q_3^2) + 
       q_{12}^2 (-q_2^2 + q_{23}^2 + q_3^2))\Bigg]\nonumber\\
&\ \hspace{-80mm} + 
 {\rm Tri}(m_0,m_1,m_2,0,q_1,q_2;D)\Bigg[m_3^2 (q_{12}^2 - q_2^2)^2 - 
    m_2^2 (q_{12}^2 - q_2^2) (q_{13}^2 - q_3^2)\nonumber\\
&\ \hspace{-70mm}
 - (q_{12}^2 - q_2^2) (-q_{13}^2 q_2^2 +
        q_{12}^2 q_3^2) + 
    m_0^2 (-q_{12}^4 + q_2^2 (q_{13}^2 - q_{23}^2)\nonumber\\
&\ \hspace{-70mm} + 
       q_{12}^2 (q_{13}^2 + q_2^2 + q_{23}^2 - 2 q_3^2)) + 
    m_1^2 (q_{12}^2 (q_2^2 - q_{23}^2 + q_3^2) + 
       q_2^2 (-2 q_{13}^2 - q_2^2 + q_{23}^2 + q_3^2))\Bigg]\nonumber\\
&\ \hspace{-80mm} + 
 {\rm Tri}(m_0,m_2,m_3,0,q_2,q_3;D)\Bigg[(q_2^2 - q_{23}^2) \Big[q_{13}^2 (m_2^2 - q_2^2) + 
       m_1^2 (q_2^2 - q_{23}^2)\Big]\nonumber\\
&\ \hspace{-70mm}
 +\Big[q_2^2 (q_{13}^2 - 2 q_{23}^2) - 
       2 m_1^2 (q_2^2 + q_{23}^2) + q_{12}^2 (q_2^2 + q_{23}^2) + 
       m_2^2 (-2 q_{12}^2 + q_{13}^2 + q_2^2 + q_{23}^2)\Big] q_3^2\nonumber\\
&\ \hspace{-70mm}
 - (-m_1^2 + m_2^2 + 
       q_{12}^2) q_3^4 + 
    m_0^2 \Big[q_{13}^2 (q_2^2 + q_{23}^2 - q_3^2) + q_{23}^2 (q_2^2 - q_{23}^2 + q_3^2) + 
       q_{12}^2 (-q_2^2 + q_{23}^2 + q_3^2)\Big]\nonumber\\
&\ \hspace{-70mm} + 
    m_3^2 (q_{12}^2 (q_2^2 - q_{23}^2 + q_3^2) + 
       q_2^2 (-2 q_{13}^2 - q_2^2 + q_{23}^2 + q_3^2))\Bigg]\nonumber\\
&\ \hspace{-80mm} + 
 {\rm Box}(m_0,m_1,m_2,m_3,0,q_1,q_2,q_3;D)\Bigg[m_3^4 (q_{12}^2 - q_2^2)^2 + 
    m_0^4 \Big[q_{12}^4 + (q_{13}^2 - q_{23}^2)^2 - 2 q_{12}^2 (q_{13}^2 +
    q_{23}^2)\Big]\nonumber\\
&\ \hspace{-70mm} + 
    m_2^4 (q_{13}^2 - q_3^2)^2 + (q_{13}^2 q_2^2 - q_{12}^2 q_3^2)^2 + 
    m_1^4 \Big[q_2^4 + (q_{23}^2 - q_3^2)^2 - 2 q_2^2 (q_{23}^2 +
    q_3^2)\Big]\nonumber\\
&\ \hspace{-70mm} + 
    2 m_1^2 \Big[q_{13}^2 q_2^2 (-q_2^2 + q_{23}^2 + q_3^2) + 
       q_3^2 (-2 q_2^2 q_{23}^2 + q_{12}^2 (q_2^2 + q_{23}^2 -
       q_3^2))\Big]\nonumber\\
&\ \hspace{-70mm} + 
    2 m_2^2 \Big[-(q_{13}^2 - q_3^2) (q_{13}^2 q_2^2 - q_{12}^2 q_3^2) + 
       m_1^2 (q_3^2 (-2 q_{12}^2 + q_2^2 + q_{23}^2 - q_3^2) + 
          q_{13}^2 (q_2^2 - q_{23}^2 + q_3^2))\Big]\nonumber\\
&\ \hspace{-70mm} + 
    2 m_0^2 \Big[q_{13}^2 q_2^2 (-q_{13}^2 + q_{23}^2) - q_{12}^4 q_3^2 + 
       m_2^2 (q_{13}^2 (q_{12}^2 - q_{13}^2 - 2 q_2^2 + q_{23}^2) + (q_{12}^2 + q_{13}^2 -
              q_{23}^2) q_3^2)\nonumber\\
&\ \hspace{-70mm} + 
       m_3^2 (-q_{12}^4 + q_2^2 (q_{13}^2 - q_{23}^2) + 
          q_{12}^2 (q_{13}^2 + q_2^2 + q_{23}^2 - 2 q_3^2)) + 
       q_{12}^2 (q_{23}^2 q_3^2 + q_{13}^2 (q_2^2 - 2 q_{23}^2 +
       q_3^2))\nonumber\\
&\ \hspace{-70mm} + 
       m_1^2 (q_{13}^2 (q_2^2 + q_{23}^2 - q_3^2) + 
          q_{23}^2 (q_2^2 - q_{23}^2 + q_3^2) + 
          q_{12}^2 (-q_2^2 + q_{23}^2 + q_3^2))\Big]\nonumber\\
&\ \hspace{-70mm} + 
    2 m_3^2 \Big[-m_2^2 (q_{12}^2 - q_2^2) (q_{13}^2 - q_3^2) - (q_{12}^2 - 
          q_2^2) (-q_{13}^2 q_2^2 + q_{12}^2 q_3^2)\nonumber\\
&\ \hspace{-70mm} + 
       m_1^2 (q_{12}^2 (q_2^2 - q_{23}^2 + q_3^2) + 
          q_2^2 (-2 q_{13}^2 - q_2^2 + q_{23}^2 + q_3^2))\Big]\Bigg] \Bigg\}\,.
\end{align}
\end{appendix}


\newpage
\begin{thebibliography}{99}
\bibitem{Group:2008vn}
  T.~T.~E.~Group, f.~t.~CDF and D.~Collaborations,
  arXiv:0808.1089 [hep-ex].

\bibitem{CMS}
CMS physics: Technical Design Report, CERN-LHCC-2006-021; CMS-TDR-008-2, {\it http://cmsdoc.cern.ch/cms/cpt/tdr/}

\bibitem{Nason:1987xz}
  P.~Nason, S.~Dawson and R.~K.~Ellis,
  Nucl.\ Phys.\  B {\bf 303}, 607 (1988).

\bibitem{Beenakker:1988bq}
  W.~Beenakker, H.~Kuijf, W.~L.~van Neerven and J.~Smith,
  Phys.\ Rev.\  D {\bf 40}, 54 (1989), \\
  W.~Beenakker, W.~L.~van Neerven, R.~Meng, G.~A.~Schuler and J.~Smith,
  Nucl.\ Phys.\  B {\bf 351}, 507 (1991).


\bibitem{Mangano:1991jk}
  M.~L.~Mangano, P.~Nason and G.~Ridolfi,
  Nucl.\ Phys.\  B {\bf 373}, 295 (1992).


\bibitem{Bernreuther:2001bx}
  W.~Bernreuther, A.~Brandenburg, Z.~G.~Si and P.~Uwer,
  Phys.\ Lett.\  B {\bf 509}, 53 (2001)
  [arXiv:hep-ph/0104096].


\bibitem{Bernreuther:2004jv}
  W.~Bernreuther, A.~Brandenburg, Z.~G.~Si and P.~Uwer,
  Nucl.\ Phys.\  B {\bf 690}, 81 (2004)
  [arXiv:hep-ph/0403035].




\bibitem{Laenen:1993xr}
  E.~Laenen, J.~Smith and W.~L.~van Neerven,
  Phys.\ Lett.\  B {\bf 321}, 254 (1994)
  [arXiv:hep-ph/9310233].


\bibitem{Berger:1996ad}
  E.~L.~Berger and H.~Contopanagos,
  Phys.\ Rev.\  D {\bf 54}, 3085 (1996)
  [arXiv:hep-ph/9603326].


\bibitem{Kidonakis:1997gm}
  N.~Kidonakis and G.~Sterman,
  Nucl.\ Phys.\  B {\bf 505}, 321 (1997)
  [arXiv:hep-ph/9705234].

\bibitem{Bonciani:1998vc}
  R.~Bonciani, S.~Catani, M.~L.~Mangano and P.~Nason,
  Nucl.\ Phys.\  B {\bf 529}, 424 (1998)
  [Erratum-ibid.\  B {\bf 803}, 234 (2008)]
  [arXiv:hep-ph/9801375].
  
\bibitem{Kidonakis:2000ui}
  N.~Kidonakis,
  Phys.\ Rev.\  D {\bf 64}, 014009 (2001)
  [arXiv:hep-ph/0010002],\\
  N.~Kidonakis, E.~Laenen, S.~Moch and R.~Vogt,
  Phys.\ Rev.\  D {\bf 64}, 114001 (2001)
  [arXiv:hep-ph/0105041],\\
  N.~Kidonakis and R.~Vogt,
  Phys.\ Rev.\  D {\bf 68}, 114014 (2003)
  [arXiv:hep-ph/0308222],\\
  N.~Kidonakis,
  Phys.\ Rev.\  D {\bf 73}, 034001 (2006)
  [arXiv:hep-ph/0509079].


\bibitem{Frixione:2003ei}
  S.~Frixione, P.~Nason and B.~R.~Webber,
  JHEP {\bf 0308}, 007 (2003)
  [arXiv:hep-ph/0305252].

\bibitem{Cacciari:2008zb}
  M.~Cacciari, S.~Frixione, M.~M.~Mangano, P.~Nason and G.~Ridolfi,
  arXiv:0804.2800 [hep-ph].

\bibitem{Moch:2008qy}
  S.~Moch and P.~Uwer,
  Phys.\ Rev.\  D {\bf 78}, 034003 (2008)
  [arXiv:0804.1476 [hep-ph]],\\
  N.~Kidonakis and R.~Vogt,
  arXiv:0805.3844 [hep-ph].


\bibitem{Hamberg:1990np}
  R.~Hamberg, W.~L.~van Neerven and T.~Matsuura,
  Nucl.\ Phys.\  B {\bf 359}, 343 (1991)
  [Erratum-ibid.\  B {\bf 644}, 403 (2002)].

\bibitem{Harlander:2002wh}
  R.~V.~Harlander and W.~B.~Kilgore,
  Phys.\ Rev.\ Lett.\  {\bf 88}, 201801 (2002)
  [arXiv:hep-ph/0201206].

\bibitem{Anastasiou:2003yy}
  C.~Anastasiou, L.~J.~Dixon, K.~Melnikov and F.~Petriello,
  Phys.\ Rev.\ Lett.\  {\bf 91}, 182002 (2003)
  [arXiv:hep-ph/0306192],\\
  C.~Anastasiou, L.~J.~Dixon, K.~Melnikov and F.~Petriello,
  Phys.\ Rev.\  D {\bf 69}, 094008 (2004)
  [arXiv:hep-ph/0312266].

\bibitem{Melnikov:2006di}
  K.~Melnikov and F.~Petriello,
  Phys.\ Rev.\ Lett.\  {\bf 96}, 231803 (2006)
  [arXiv:hep-ph/0603182],\\
  K.~Melnikov and F.~Petriello,
  Phys.\ Rev.\  D {\bf 74}, 114017 (2006)
  [arXiv:hep-ph/0609070].


\bibitem{Anastasiou:2002yz}
  C.~Anastasiou and K.~Melnikov,
  Nucl.\ Phys.\  B {\bf 646}, 220 (2002)
  [arXiv:hep-ph/0207004].


\bibitem{Ravindran:2003um}
  V.~Ravindran, J.~Smith and W.~L.~van Neerven,
  Nucl.\ Phys.\  B {\bf 665}, 325 (2003)
  [arXiv:hep-ph/0302135].

\bibitem{Anastasiou:2004xq}
  C.~Anastasiou, K.~Melnikov and F.~Petriello,
  Phys.\ Rev.\ Lett.\  {\bf 93}, 262002 (2004)
  [arXiv:hep-ph/0409088].

\bibitem{Anastasiou:2005qj}
  C.~Anastasiou, K.~Melnikov and F.~Petriello,
  Nucl.\ Phys.\  B {\bf 724}, 197 (2005)
  [arXiv:hep-ph/0501130].

\bibitem{Catani:2007vq}
  S.~Catani and M.~Grazzini,
  Phys.\ Rev.\ Lett.\  {\bf 98}, 222002 (2007)
  [arXiv:hep-ph/0703012].

\bibitem{Anastasiou:2007mz}
  C.~Anastasiou, G.~Dissertori and F.~Stockli,
  JHEP {\bf 0709}, 018 (2007)
  [arXiv:0707.2373 [hep-ph]].

\bibitem{Grazzini:2008tf}
  M.~Grazzini,
  JHEP {\bf 0802}, 043 (2008)
  [arXiv:0801.3232 [hep-ph]].

\bibitem{Anastasiou:2008ik}
  C.~Anastasiou, G.~Dissertori, F.~Stockli and B.~R.~Webber,
  JHEP {\bf 0803}, 017 (2008)
  [arXiv:0801.2682 [hep-ph]].

\bibitem{DavatzCMSnote}
G. Davatz, M. Dittmar, A.-S. Giolo-Nicollerat, CMS Note 2006/047,\\
G.~Davatz, M.~Dittmar and F.~Pauss,
arXiv:hep-ph/0612099,\\
G.~Davatz, M.~Dittmar and A.~S.~Giolo-Nicollerat,
J.\ Phys.\ G {\bf 33}, N85 (2007),\\
G.~Davatz, A.~S.~Giolo-Nicollerat and M.~Zanetti,CERN-CMS-NOTE-2006-048.



\bibitem{Campbell:2006wx}
  J.~M.~Campbell, J.~W.~Huston and W.~J.~Stirling,
  Rept.\ Prog.\ Phys.\  {\bf 70}, 89 (2007)
  [arXiv:hep-ph/0611148].

\bibitem{Czakon:2007ej}
  M.~Czakon, A.~Mitov and S.~Moch,
  Phys.\ Lett.\  B {\bf 651}, 147 (2007)
  [arXiv:0705.1975 [hep-ph]].



\bibitem{Czakon:2007wk}
  M.~Czakon, A.~Mitov and S.~Moch,
  Nucl.\ Phys.\  B {\bf 798}, 210 (2008)
  [arXiv:0707.4139 [hep-ph]].

\bibitem{Mitov:2006xs}
  A.~Mitov and S.~Moch,
  JHEP {\bf 0705}, 001 (2007)
  [arXiv:hep-ph/0612149],\\
  T.~Becher and K.~Melnikov,
  JHEP {\bf 0706}, 084 (2007)
  [arXiv:0704.3582 [hep-ph]].

\bibitem{Anastasiou:2000kg}
  C.~Anastasiou, E.~W.~N.~Glover, C.~Oleari and M.~E.~Tejeda-Yeomans,
  Nucl.\ Phys.\  B {\bf 605}, 486 (2001)
  [arXiv:hep-ph/0101304], \\
  C.~Anastasiou, E.~W.~N.~Glover, C.~Oleari and M.~E.~Tejeda-Yeomans,
  Phys.\ Lett.\  B {\bf 506}, 59 (2001)
  [arXiv:hep-ph/0012007], \\
  C.~Anastasiou, E.~W.~N.~Glover, C.~Oleari and M.~E.~Tejeda-Yeomans,
  Nucl.\ Phys.\  B {\bf 601}, 318 (2001)
  [arXiv:hep-ph/0010212],\\
  Z.~Bern, A.~De Freitas and L.~J.~Dixon,
  JHEP {\bf 0306}, 028 (2003)
  [arXiv:hep-ph/0304168],\\
  E.~W.~N.~Glover and M.~E.~Tejeda-Yeomans,
  JHEP {\bf 0306}, 033 (2003)
  [arXiv:hep-ph/0304169],\\
  E.~W.~N.~Glover,
  JHEP {\bf 0404}, 021 (2004)
  [arXiv:hep-ph/0401119],\\
  A.~De Freitas and Z.~Bern,
  JHEP {\bf 0409}, 039 (2004)
  [arXiv:hep-ph/0409007].






\bibitem{Czakon:2008zk}
  M.~Czakon,
  Phys.\ Lett.\  B {\bf 664}, 307 (2008)
  [arXiv:0803.1400 [hep-ph]].

\bibitem{Bonciani:2008az}
  R.~Bonciani, A.~Ferroglia, T.~Gehrmann, D.~Maitre and C.~Studerus,
  arXiv:0806.2301 [hep-ph].

\bibitem{Dittmaier:2007wz}
  S.~Dittmaier, P.~Uwer and S.~Weinzierl,
  Phys.\ Rev.\ Lett.\  {\bf 98}, 262002 (2007)
  [arXiv:hep-ph/0703120].

\bibitem{GehrmannDeRidder:2007jk}
  A.~Gehrmann-De Ridder, T.~Gehrmann, E.~W.~N.~Glover and G.~Heinrich,
  JHEP {\bf 0711}, 058 (2007)
  [arXiv:0710.0346 [hep-ph]],\\
  A.~Gehrmann-De Ridder, T.~Gehrmann, E.~W.~N.~Glover and G.~Heinrich,
  JHEP {\bf 0712}, 094 (2007)
  [arXiv:0711.4711 [hep-ph]],\\
  A.~Gehrmann-De Ridder, T.~Gehrmann, E.~W.~N.~Glover and G.~Heinrich,
  Phys.\ Rev.\ Lett.\  {\bf 100}, 172001 (2008)
  [arXiv:0802.0813 [hep-ph]],\\
  S.~Weinzierl,
  arXiv:0807.3241 [hep-ph].


\bibitem{Anastasiou:2004qd}
  C.~Anastasiou, K.~Melnikov and F.~Petriello,
  Phys.\ Rev.\ Lett.\  {\bf 93}, 032002 (2004)
  [arXiv:hep-ph/0402280],\\
  C.~Anastasiou, K.~Melnikov and F.~Petriello,
  JHEP {\bf 0709}, 014 (2007)
  [arXiv:hep-ph/0505069].

\bibitem{Korner:2002hy}
  J.~G.~Korner and Z.~Merebashvili,
  Phys.\ Rev.\  D {\bf 66}, 054023 (2002)
  [arXiv:hep-ph/0207054],\\
  J.~G.~Korner, Z.~Merebashvili and M.~Rogal,
  Phys.\ Rev.\  D {\bf 73}, 034030 (2006)
  [arXiv:hep-ph/0511264].

\bibitem{Korner:2008bn}
  J.~G.~Korner, Z.~Merebashvili and M.~Rogal,
  Phys.\ Rev.\  D {\bf 77}, 094011 (2008)
  [arXiv:0802.0106 [hep-ph]].

\bibitem{Ossola:2006us}
  G.~Ossola, C.~G.~Papadopoulos and R.~Pittau,
  Nucl.\ Phys.\  B {\bf 763}, 147 (2007)
  [arXiv:hep-ph/0609007].

\bibitem{Ossola:2007bb}
  G.~Ossola, C.~G.~Papadopoulos and R.~Pittau,
  JHEP {\bf 0707}, 085 (2007)
  [arXiv:0704.1271 [hep-ph]].


\bibitem{Ossola:2008xq}
  G.~Ossola, C.~G.~Papadopoulos and R.~Pittau,
  JHEP {\bf 0805}, 004 (2008)
  [arXiv:0802.1876 [hep-ph]].

\bibitem{delAguila:2004nf}
  F.~del Aguila and R.~Pittau,
  JHEP {\bf 0407}, 017 (2004)
  [arXiv:hep-ph/0404120].

\bibitem{Ellis:2007br}
  R.~K.~Ellis, W.~T.~Giele and Z.~Kunszt,
  JHEP {\bf 0803}, 003 (2008)
  [arXiv:0708.2398 [hep-ph]].

\bibitem{Giele:2008ve}
  W.~T.~Giele, Z.~Kunszt and K.~Melnikov,
  JHEP {\bf 0804}, 049 (2008)
  [arXiv:0801.2237 [hep-ph]].


\bibitem{Berger:2008sj}
  C.~F.~Berger {\it et al.},
  Phys.\ Rev.\  D {\bf 78}, 036003 (2008)
  [arXiv:0803.4180 [hep-ph]].

\bibitem{Ellis:2007qk}
  R.~K.~Ellis and G.~Zanderighi,
  JHEP {\bf 0802}, 002 (2008)
  [arXiv:0712.1851 [hep-ph]].
\bibitem{Giele:2008bc}
  W.~T.~Giele and G.~Zanderighi,
  arXiv:0805.2152 [hep-ph].



\bibitem{Binoth:2008kt}
  T.~Binoth, G.~Ossola, C.~G.~Papadopoulos and R.~Pittau,
  JHEP {\bf 0806}, 082 (2008)
  [arXiv:0804.0350 [hep-ph]].

\bibitem{Bredenstein:2008zb}
  A.~Bredenstein, A.~Denner, S.~Dittmaier and S.~Pozzorini,
  arXiv:0807.1248 [hep-ph],\\
  A.~Bredenstein, A.~Denner, S.~Dittmaier and M.~M.~Weber,
  JHEP {\bf 0702}, 080 (2007)
  [arXiv:hep-ph/0611234],\\
  A.~Bredenstein, A.~Denner, S.~Dittmaier and M.~M.~Weber,
  Nucl.\ Phys.\ Proc.\ Suppl.\  {\bf 160}, 131 (2006)
  [arXiv:hep-ph/0607060],\\
  A.~Denner, S.~Dittmaier, M.~Roth and L.~H.~Wieders,
  Nucl.\ Phys.\  B {\bf 724}, 247 (2005)
  [arXiv:hep-ph/0505042],\\
  A.~Denner, S.~Dittmaier, M.~Roth and L.~H.~Wieders,
  Phys.\ Lett.\  B {\bf 612}, 223 (2005)
  [arXiv:hep-ph/0502063],\\
  A.~Denner, S.~Dittmaier, M.~Roth and M.~M.~Weber,
  Phys.\ Lett.\  B {\bf 575}, 290 (2003)
  [arXiv:hep-ph/0307193],\\
  A.~Denner, S.~Dittmaier, M.~Roth and M.~M.~Weber,
  Nucl.\ Phys.\  B {\bf 660}, 289 (2003)
  [arXiv:hep-ph/0302198].


\bibitem{Lazopoulos:2008de}
  A.~Lazopoulos, T.~McElmurry, K.~Melnikov and F.~Petriello,
  Phys.\ Lett.\  B {\bf 666}, 62 (2008)
  [arXiv:0804.2220 [hep-ph]],\\
  A.~Lazopoulos, K.~Melnikov and F.~J.~Petriello,
  Phys.\ Rev.\  D {\bf 77}, 034021 (2008)
  [arXiv:0709.4044 [hep-ph]],\\
  A.~Lazopoulos, K.~Melnikov and F.~Petriello,
  Phys.\ Rev.\  D {\bf 76}, 014001 (2007)
  [arXiv:hep-ph/0703273].

\bibitem{Giele:2004iy}
  W.~T.~Giele and E.~W.~N.~Glover,
  JHEP {\bf 0404}, 029 (2004)
  [arXiv:hep-ph/0402152].

\bibitem{Campbell:2007ev}
  J.~M.~Campbell, R.~Keith Ellis and G.~Zanderighi,
  JHEP {\bf 0712}, 056 (2007)
  [arXiv:0710.1832 [hep-ph]],\\
  J.~M.~Campbell, R.~K.~Ellis and G.~Zanderighi,
  JHEP {\bf 0610}, 028 (2006)
  [arXiv:hep-ph/0608194],

\bibitem{Campanario:2008yg}
  F.~Campanario, V.~Hankele, C.~Oleari, S.~Prestel and D.~Zeppenfeld,
  arXiv:0809.0790 [hep-ph],\\
  G.~Bozzi, B.~Jager, C.~Oleari and D.~Zeppenfeld,
  Phys.\ Rev.\  D {\bf 75}, 073004 (2007)
  [arXiv:hep-ph/0701105],\\
  B.~Jager, C.~Oleari and D.~Zeppenfeld,
  JHEP {\bf 0607}, 015 (2006)
  [arXiv:hep-ph/0603177].


\bibitem{Davydychev:1991va}
  A.~I.~Davydychev,
  Phys.\ Lett.\  B {\bf 263}, 107 (1991).


\bibitem{Anastasiou:1999bn}
  C.~Anastasiou, E.~W.~N.~Glover and C.~Oleari,
  Nucl.\ Phys.\  B {\bf 575}, 416 (2000)
  [Erratum-ibid.\  B {\bf 585}, 763 (2000)]
  [arXiv:hep-ph/9912251].



\bibitem{Anastasiou:2006jv}
  C.~Anastasiou, R.~Britto, B.~Feng, Z.~Kunszt and P.~Mastrolia,
  Phys.\ Lett.\  B {\bf 645}, 213 (2007)
  [arXiv:hep-ph/0609191],\\
  C.~Anastasiou, R.~Britto, B.~Feng, Z.~Kunszt and P.~Mastrolia,
  JHEP {\bf 0703}, 111 (2007)
  [arXiv:hep-ph/0612277].

\bibitem{Ellis:2008ir}
  R.~K.~Ellis, W.~T.~Giele, Z.~Kunszt and K.~Melnikov,
  arXiv:0806.3467 [hep-ph].

\bibitem{unitarity}
R.~E.~Cutkosky, J. Math. Phys. 1, 429 (1960),\\
G.~'t~Hooft and M.~Veltman, Diagrammar, CERN Report 73-9, Geneva (1973),\\
  W.~L.~van Neerven,
  Nucl.\ Phys.\ B 268, 453 (1986).

\bibitem{Bern:1994zx}
  Z.~Bern, L.~J.~Dixon, D.~C.~Dunbar and D.~A.~Kosower,
  Nucl.\ Phys.\ B 425, 217 (1994) [hep-ph/9403226], \\
Z.~Bern, L.~J.~Dixon, D.~C.~Dunbar and D.~A.~Kosower,
Nucl.\ Phys.\ B 435, 59 (1995)
[hep-ph/9409265].

\bibitem{Bern:1995db}
  Z.~Bern and A.~G.~Morgan,
  Nucl.\ Phys.\  B {\bf 467}, 479 (1996)
  [arXiv:hep-ph/9511336].

\bibitem{Bern:1996}
  Z.~Bern, L.~J.~Dixon and D.~A.~Kosower,
  Ann.\ Rev.\ Nucl.\ Part.\ Sci.\  {\bf 46}, 109 (1996)
  [arXiv:hep-ph/9602280],\\
  Z.~Bern, L.~J.~Dixon, D.~C.~Dunbar and D.~A.~Kosower,
  Phys.\ Lett.\  B {\bf 394}, 105 (1997)
  [arXiv:hep-th/9611127].

\bibitem{Bern:1997sc}
  Z.~Bern, L.~J.~Dixon and D.~A.~Kosower,
  Nucl.\ Phys.\  B {\bf 513}, 3 (1998)
  [arXiv:hep-ph/9708239].


\bibitem{BCFgeneralized}
R.~Britto, F.~Cachazo and B.~Feng,
Nucl.\ Phys.\  B 725, 275 (2005)
[hep-th/0412103].

\bibitem{Bern:2002zk}
  Z.~Bern, A.~De Freitas, L.~J.~Dixon and H.~L.~Wong,
  Phys.\ Rev.\  D {\bf 66}, 085002 (2002)
  [arXiv:hep-ph/0202271].

\bibitem{Anastasiou:2008rm}
  C.~Anastasiou, S.~Beerli and A.~Daleo,
  Phys.\ Rev.\ Lett.\  {\bf 100}, 241806 (2008)
  [arXiv:0803.3065 [hep-ph]].

\bibitem{Harlander:2006rj}
  R.~Harlander, P.~Kant, L.~Mihaila and M.~Steinhauser,
  JHEP {\bf 0609}, 053 (2006)
  [arXiv:hep-ph/0607240].

\bibitem{'tHooft:1972fi}
  G.~'t Hooft and M.~J.~G.~Veltman,
  Nucl.\ Phys.\  B {\bf 44}, 189 (1972).

\bibitem{Garland:2002ak}
  L.~W.~Garland, T.~Gehrmann, E.~W.~N.~Glover, A.~Koukoutsakis and E.~Remiddi,
  Nucl.\ Phys.\  B {\bf 642}, 227 (2002)
  [arXiv:hep-ph/0206067].



\bibitem{Nogueira:1991ex}
  P.~Nogueira,
  J.\ Comput.\ Phys.\  {\bf 105}, 279 (1993).

\bibitem{Vermaseren:2000nd}
  J.~A.~M.~Vermaseren,
  arXiv:math-ph/0010025.

\bibitem{Anastasiou:2004vj}
  C.~Anastasiou and A.~Lazopoulos,
  JHEP {\bf 0407}, 046 (2004)
  [arXiv:hep-ph/0404258].

\bibitem{Bern:1993kr}
  Z.~Bern, L.~J.~Dixon and D.~A.~Kosower,
  Nucl.\ Phys.\  B {\bf 412}, 751 (1994)
  [arXiv:hep-ph/9306240].

\bibitem{Campbell:1996zw}
  J.~M.~Campbell, E.~W.~N.~Glover and D.~J.~Miller,
  Nucl.\ Phys.\  B {\bf 498}, 397 (1997)
  [arXiv:hep-ph/9612413].


\bibitem{Britto:2008vq}
  R.~Britto, B.~Feng and P.~Mastrolia,
  Phys.\ Rev.\  D {\bf 78}, 025031 (2008)
  [arXiv:0803.1989 [hep-ph]].



\bibitem{Britto:2006sj}
  R.~Britto, B.~Feng and P.~Mastrolia,
  Phys.\ Rev.\  D {\bf 73}, 105004 (2006)
  [arXiv:hep-ph/0602178].

\bibitem{Feng:2008ju}
  B.~Feng and G.~Yang,
  arXiv:0806.4016 [hep-ph].


\bibitem{Britto:2008sw}
  R.~Britto, B.~Feng and G.~Yang,
  arXiv:0803.3147 [hep-ph].

\bibitem{Britto:2007tt}
  R.~Britto and B.~Feng,
  JHEP {\bf 0802}, 095 (2008)
  [arXiv:0711.4284 [hep-ph]].

\bibitem{Anastasiou:2006gt}
  C.~Anastasiou, R.~Britto, B.~Feng, Z.~Kunszt and P.~Mastrolia,
  JHEP {\bf 0703}, 111 (2007)
  [arXiv:hep-ph/0612277].

\bibitem{Korner:2004im}
  J.~G.~Korner, Z.~Merebashvili and M.~Rogal,
  Nucl.\ Phys.\ Proc.\ Suppl.\  {\bf 135}, 285 (2004)
  [arXiv:hep-ph/0411394].

\bibitem{Catani:2000ef}
  S.~Catani, S.~Dittmaier and Z.~Trocsanyi,
  Phys.\ Lett.\  B {\bf 500}, 149 (2001)
  [arXiv:hep-ph/0011222].
\end{thebibliography}
\end{document}